\documentclass[lettersize,journal]{IEEEtran}
\usepackage{graphicx}
\usepackage[most]{tcolorbox} 
\usepackage{tcolorbox}
\usepackage{caption}
\usepackage{booktabs}
\usepackage{amssymb}
\usepackage{amsmath}
\usepackage{algorithm}
\usepackage{algpseudocode}
\usepackage{tcolorbox}
\usepackage{booktabs}
\usepackage{float}
\usepackage{listings}
\usepackage{tabularx}
\usepackage{hyperref}

\usepackage[most]{tcolorbox}
\usepackage{array}
\usepackage{siunitx}
\usepackage{graphicx}
\usepackage{tabularx}
\usepackage{xcolor}
\definecolor{logbg}{RGB}{250,250,250}
\definecolor{logborder}{RGB}{180,180,180}
\definecolor{logtext}{RGB}{30,30,30}
\definecolor{logkey}{RGB}{0,100,180}
\definecolor{logvalue}{RGB}{180,30,30}
\lstdefinestyle{logstyle}{
  basicstyle=\ttfamily\footnotesize\color{logtext},
  backgroundcolor=\color{logbg},
  frame=single,
  rulecolor=\color{logborder},
  xleftmargin=1em,
  breaklines=true,
  keywordstyle=\color{logkey},
  stringstyle=\color{logvalue},
  showstringspaces=false
}
\usepackage{graphicx}
\def\BibTeX{{\rm B\kern-.05em{\sc i\kern-.025em b}\kern-.08em
    T\kern-.1667em\lower.7ex\hbox{E}\kern-.125emX}}
\usepackage{balance}

\begin{document}
\title{Think Fast: Real-Time IoT Intrusion Reasoning Using IDS and LLMs at the Edge Gateway}

\author{
Saeid Jamshidi\thanks{Saeid Jamshidi, Omar Abdul Wahab, and Martine Bellaïche are with the Department of Computer and Software Engineering, Polytechnique Montréal, Quebec, H3T 1J4, Canada. Contact emails: jamshidi.saeid@polymtl.ca, omar.abdul-wahab@polymtl.ca, martine.bellaiche@polymtl.ca.}
\and
Amin Nikanjam\thanks{Amin Nikanjam was with the Huawei Distributed Scheduling and Data Engine Lab (work done while at Polytechnique Montréal), Montreal, Quebec, Canada. Email: amin.nikanjam@huawei.com.}
\and
Negar Shahabi\thanks{Negar Shahabi is with the Concordia Institute for Information Systems Engineering (CIISE), Concordia University, Montréal, Quebec, Canada. Email: negar.shahabi@mail.concordia.ca.}
\and
Kawser Wazed Nafi\thanks{Kawser Wazed Nafi and Foutse Khomh are with the SWAT Laboratory, Polytechnique Montréal, Quebec, H3T 1J4, Canada. Emails: kawser.wazed-nafi@polymtl.ca, foutse.khomh@polymtl.ca.}
\and
Foutse Khomh
\and
Samira Keivanpour\thanks{Samira Keivanpour is with Poly Circle X.O, Polytechnique Montréal, Quebec, H3T 1J4, Canada. Email: samira.keivanpour@polymtl.ca.}
\and
Rolando Herrero\thanks{Rolando Herrero is with the College of Engineering, Northeastern University, Boston, MA, USA. Email: r.herrero@northeastern.edu.}
\and
Omar Abdul Wahab
\and
Martine Bellaïche
}

\maketitle

\begin{abstract}
As the number of connected IoT devices continues to grow, securing these systems against cyber threats remains a pressing challenge, especially within environments constrained by limited computational and energy resources. This paper presents an edge-centric Intrusion Detection System (IDS) framework that seamlessly integrates lightweight Machine Learning (ML)-based IDS models with pre-trained Large Language Models (LLMs) to enhance detection accuracy, semantic interpretability, and operational efficiency at the network edge. The system evaluates six ML-based IDS models: Decision Tree (DT), K-Nearest Neighbors (KNN), Random Forest (RF), Convolutional Neural Network (CNN), Long Short-Term Memory (LSTM), and a hybrid model of CNN and LSTM, on low-power edge gateways, achieving accuracy up to 98\% under real-world cyberattacks. Furthermore, for anomaly detection, the system transmits a compact, secure telemetry snapshot (e.g., CPU usage, memory usage, latency, and energy consumption) via low-bandwidth API calls to LLMs, including GPT-4-turbo, DeepSeek V2, and LLaMA 3.5. These models employ zero-shot, few-shot, and Chain-of-Thought (CoT) reasoning to deliver human-readable threat analyses and actionable mitigation recommendations. Extensive evaluations across diverse attacks such as DoS, DDoS, brute force and port scanning,  demonstrate the system’s ability to enhance interpretability while maintaining low latency ($<$1.5s), minimal bandwidth usage ($<$1.2 kB per prompt), and energy efficiency ($<$75 J), establishing it as a practical and scalable IDS solution for the edge gateway.
\end{abstract}


\maketitle

\section{Introduction}
\label{Introduction}
The Internet of Things (IoT) has become a fundamental layer of modern digital infrastructure\cite{vermesan2022digitising}, connecting billions of physical devices that range from industrial sensors to home automation systems \cite{lampropoulos2019internet} \cite{perwej2019internet}. This massive integration enables real-time control, predictive analytics, and pervasive automation in various domains, including healthcare, energy, agriculture, and transportation \cite{holler2014internet}. However, as this connectivity becomes more ubiquitous, it also brings with it a rapidly expanding and deeply fragmented attack surface \cite{butun2019security}. Most IoT devices operate with limited processing power, minimal memory, and very modest energy budgets \cite{capra2019edge}. They are often deployed with weak default credentials, lack regular firmware updates, and run on lightweight operating systems with minimal security controls \cite{yang2017hardware} \cite{akmandor2018smart}. As a result, they are particularly vulnerable to attacks, e.g., Distributed Denial-of-Service (DDoS), brute-force attacks, and port scanning \cite{jamshidi2025evaluating}. These threats are not just technical nuisances; they can disrupt safety-critical processes or silently compromise sensitive infrastructure.\\
Furthermore, what makes securing IoT networks challenging is the tight coupling between performance and resource constraints. Many devices must operate with strict limitations on energy consumption and CPU usage \cite{nilima2024optimizing} \cite{aziz2023securing}. A slight increase in computational load and energy consumption can cause an entire sensor cluster to fail prematurely, missing critical timing windows \cite{darabkh2025evolutionary}. Therefore, any effective security solution for the IoT must strike a balance between detection effectiveness and computational efficiency. It must be able to accurately identify threats while preserving the functional integrity of the device and respecting its operational boundaries \cite{heidari2023internet}.\\
One approach that has shown promise is the use of ML-based IDS at the edge gateway \cite{gyamfi2022intrusion, manivannan2024recent}. These systems learn the expected behavior of a device and network and flag deviations without relying on static signatures—predefined patterns of known threats stored in a database that traditional signature-based IDS use for detection. By avoiding dependence on such fixed patterns, ML-based IDS can identify both known and previously unseen attacks. When properly tuned, these models can achieve high detection accuracy with minimal supervision \cite{sahani2023machine}. Moreover, their lightweight nature makes them suitable for on-device deployment, reducing reliance on external infrastructure and enabling real-time detection \cite{amgbara2024exploring}. However, ML-based IDS models also face persistent challenges. They tend to produce a high rate of false positives, which undermines trust and usability. They often return binary and numeric results without contextual explanation, leaving humans in the loop to infer the meaning of each alert \cite{duraz2024trustable} \cite{afifi2024machine}. Additionally, when these models are deployed on resource-limited platforms, e.g., single-board computers, they must be simplified, which can result in a loss of precision and reliability \cite{dina2021intrusion}.\\
Concurrently, large language models (LLMs), e.g., GPT-4-turbo, DeepSeek-V2, and Llama3.5, have emerged as robust mechanisms for semantic reasoning and pattern interpretation. They can identify complex relationships in data and generate coherent explanations, even from limited examples, a capability often referred to as few-shot learning \cite{lucafuture}. These models demonstrate significant strength in interpreting intricate inputs and producing human-readable insights \cite{bhat2023human}\cite{agarwal2025review}.
In particular, their ability to generalize from limited examples and perform reasoning tasks across structured and unstructured datasets provides promising avenues for enhancing security applications. Nevertheless, the integration of LLMs within ML-based IDS  pipelines introduces considerable practical challenges \cite{sarhaddi2025llms}. Key concerns include evaluating whether the LLM can effectively augment ML-based IDS outputs without substantially increasing computational load and breaching latency and energy consumption constraints. Additionally, it remains crucial to determine whether pre-trained LLMs can accurately interpret compact telemetry from the edge gateway and whether such interpretations can significantly enhance detection accuracy under realistic attack conditions.\\
To address these challenges, this paper proposes a novel edge-centric IDS architecture, where six ML-based IDS models are deployed exclusively on a low-power edge gateway. Upon detecting an anomaly, the gateway compiles a concise snapshot of key system metrics, e.g., CPU usage, memory usage, latency, bandwidth, and energy consumption. This aggregated telemetry is then transmitted to large LLMs via encrypted, low-overhead API calls. Each LLM performs a structured, multi-stage analysis encompassing zero-shot classification to hypothesize the nature of the attack, few-shot reasoning to assess pattern similarity, and Chain-of-Thought (CoT) reasoning to produce contextually relevant explanations and potential countermeasures. The resulting alert is semantically enhanced, significantly improving interpretability and facilitating more informed decision-making. Importantly, all processes, e.g., orchestration, metric aggregation, and integration, are executed directly on the edge gateway. The significant contributions of this paper can be summarized as follows:
\begin{itemize}
\item \textbf{Edge-Initiated Semantic Defence}: A fully autonomous ML-based IDS architecture where detection and coordination occur entirely on the edge gateway, leveraging external LLMs solely through minimal API calls to enhance reasoning capabilities.
\item \textbf{Multi-Stage LLM Reasoning Integration}: Integration of advanced LLM reasoning methods, zero-shot classification, few-shot reasoning, and CoT processing, to convert raw anomaly alerts into structured, interpretable, and actionable security intelligence.
\item \textbf{Resource-Conscious Evaluation Framework}: An evaluation approach that simultaneously assesses detection accuracy and precisely measures system resource impacts,e.g, latency, CPU usage, memory usage, and energy consumption, thus balancing effectiveness with practical feasibility.
\end{itemize}

The remainder of this paper is organized as follows. Section~\ref{Background} introduces the foundational ML-based IDS. Section~\ref{relatedwork} reviews prior research on ML-based IDS and LLM integration for IoT security. Section~\ref{sec:proposed_method} details the proposed system, covering the inference pipeline, telemetry, context augmentation, prompt construction, reasoning strategies, and evaluation framework. Section~\ref{Study design} describes the research questions, datasets, and experimental setup. Section~\ref{Experimental results and analysis} presents the experimental outcomes across various attack scenarios and resource metrics. Section~\ref{Discussion} discusses key findings, model performance, and interpretability improvements. Section~\ref{sec:T_V} outlines threats to validity that impact the study's generalizability. Section~\ref{Future Work} suggests future enhancements, including retraining, LLM optimization, and real-world deployment. Section~\ref{Conclusion} concludes the paper by summarizing the contributions and emphasizing the trade-off between detection accuracy, interpretability, and efficiency.

\section{Background}
\label{Background}
This section introduces the baseline ML-based models underpinning our research and their relevance to IDS. Each model was selected for its strengths in resource efficiency, detection accuracy, and interpretability. Together, they form the foundation upon which LLM-based semantic reasoning is integrated.\\  
\textbf{Decision Tree (DT):} DTs are widely deployed in IDS due to their interpretable, rule-based structure. By recursively partitioning traffic into binary splits, they generate hierarchical paths that clearly map input features to attack decisions. Their transparency and low computational overhead make them well-suited for resource-constrained IoT environments and real-time contexts \cite{ingre2017decision, azam2023comparative, bouke2022e2ids}.\\  
\textbf{Random Forest (RF):} RF improves upon DT by combining multiple decision trees into an ensemble. By leveraging majority voting across trees, RF reduces variance and minimizes overfitting, resulting in robust classification even with high-dimensional or imbalanced IoT traffic \cite{wali2025explainable, choubisa2022simple, markovic2022random}.\\
\textbf{K-Nearest Neighbor (KNN):} KNN performs instance-based classification by comparing feature vectors with labeled examples using distance metrics. Its non-parametric nature and adaptability make it effective for anomaly detection tasks in IDS, especially when training resources are limited \cite{mohy2023efficient, ozturk2023high, mohy2023intrusion}.\\
\textbf{Long Short-Term Memory (LSTM):} LSTMs, a variant of recurrent neural networks (RNNs), are designed to model sequential dependencies. Their gated memory cells capture long-range temporal patterns, making them highly effective for detecting evolving attacks such as slow brute force attempts and distributed DoS floods \cite{laghrissi2021intrusion, awad2023improved}.\\  
\textbf{Convolutional Neural Network (CNN):} CNNs extract spatial correlations in traffic by applying convolutional filters to feature maps. In IDS, CNNs excel at detecting both well-known and novel attack vectors by capturing structural regularities in traffic flow, including packet bursts or scanning patterns \cite{halbouni2022cnn, mohammadpour2022survey}.\\  
\textbf{Hybrid of model CNN and LSTM:} The hybrid approach integrates CNN’s spatial extraction with LSTM’s temporal reasoning, enabling richer multi-perspective analysis. Such hybrids are particularly effective for complex, multi-stage attacks that require recognition of both packet-level bursts and long-term behavioral deviations \cite{abdulmajeed2022mlids22, halbouni2022cnn}.  

\subsection{The ML-based IDS}
\label{The ML-based IDS}
We benchmarked six ML-based IDS models( e.g., DT, RF, KNN, LSTM, CNN, and a hybrid model of CNN and LSTM) under realistic edge gateway conditions to establish a baseline for efficiency and effectiveness. Each model was optimized for edge deployment and evaluated against system-level performance metrics, including CPU usage, memory usage, latency, bandwidth, and energy consumption. Building on these results, this study integrates LLM-based semantic reasoning to enhance interpretability and resilience in IDS. \\
A comparative summary of neural network–based IDS configurations is presented in Table~\ref{1_M}. The results highlight the trade-off between structural complexity and accuracy. For example, the hybrid model of CNN and LSTM  achieved the highest testing accuracy (95.75\%) while requiring 12,795 parameters, compared to the CNN model, which achieved 94.74\% testing accuracy with only 3,497 parameters. This demonstrates that although hybrid architectures provide stronger adaptability to complex traffic patterns, they impose a higher computational burden that may limit suitability for resource-constrained IoT gateways.\\  
\begin{table*}[h]
\centering
\caption{Comparison of Neural Network Models}
\label{1_M}
\renewcommand{\arraystretch}{1.3}
\begin{tabular}{|l|c|c|c|}
\hline
\textbf{Metric} & \textbf{LSTM} & \textbf{CNN + LSTM} & \textbf{CNN} \\
\hline
Dataset & CICIDS2017 & CICIDS2017 & CICIDS2017 \\
\hline
Number of Categories & 15 & 15 & 15 \\
\hline
Number of Layers & 10 & 11 & 8 \\
\hline
Number of Parameters & 56,386 & 12,795 & 3,497 \\
\hline
Training Accuracy & 97.72\% & 98.77\% & 97.92\% \\
\hline
Testing Accuracy & 93.86\% & 95.75\% & 94.74\% \\
\hline
\end{tabular}
\end{table*}
Table~\ref{table:ml_performance_comparison} further compares classical ML-based IDS models (DT, KNN, RF) against neural networks (LSTM, CNN, a hybrid model of CNN and LSTM). Classical ML-based IDS models consistently achieved near-perfect scores across accuracy, precision, recall, and F1 (99\%), confirming their efficiency and suitability for lightweight IoT deployments. Neural models, while slightly less efficient, offered improved generalization for complex and evolving attack patterns, but at the cost of increased latency and higher energy demand. This trade-off underscores the motivation for augmenting IDS with LLM-based reasoning, where semantic interpretability can alleviate computational constraints and improve robustness against sophisticated attack scenarios.  
\begin{table*}[t]
\centering
\caption{Performance Comparison of ML-based IDS Models}
\label{table:ml_performance_comparison}
\renewcommand{\arraystretch}{1.3}
\begin{tabular}{|l|c|c|c|c|c|c|}
\hline
\textbf{Metric} & \textbf{DT} & \textbf{KNN} & \textbf{RF} & \textbf{LSTM} & \textbf{LSTM + CNN} & \textbf{CNN} \\
\hline
Accuracy  & 0.9985 & 0.9967 & 0.9981 & 0.9386 & 0.9575 & 0.9474 \\
\hline
Precision & 0.9985 & 0.9966 & 0.9980 & 0.9771 & 0.9877 & 0.9792 \\
\hline
Recall    & 0.9985 & 0.9967 & 0.9981 & 0.9524 & 0.9645 & 0.9611 \\
\hline
F1-Score  & 0.9985 & 0.9966 & 0.9980 & 0.9646 & 0.9760 & 0.9701 \\
\hline
\end{tabular}
\end{table*}
\subsection{Evaluation Metrics}
\label{sec:metrics}
To assess feasibility in edge-deployed environments, we evaluate the proposed ML-based IDS and LLM hybrids using six system-level performance metrics: CPU usage, memory usage, energy consumption, bandwidth, latency, and detection accuracy. These metrics together capture computational demand, communication overhead, and energy efficiency, all of which are critical in resource-constrained edge gateways.\\
CPU usage and memory usage quantify the computational complexity of each ML-based IDS model, directly reflecting its scalability for low-power hardware. Energy consumption is measured through fine-grained integration of instantaneous power over inference cycles:
\begin{equation}
E_t = \sum_{i=1}^{N} P_i \cdot \Delta\tau, 
\quad \Delta\tau = 10\,\text{ms}, 
\quad N = \frac{T_t}{\Delta\tau}
\end{equation}
where $P_i$ denotes instantaneous power, ensuring resilience evaluation against adversarial energy-drain scenarios. Latency is modeled as the total detection delay:
\begin{equation}
T_t = T_{\text{IDS}} + T_{\text{tx}} + T_{\text{LLM}}
\end{equation}
where $T_{\text{IDS}}$ represents local inference time, $T_{\text{tx}}$ network delay, and $T_{\text{LLM}}$ reasoning time. This guarantees the IDS can be validated against real-time operational constraints.\\
Bandwidth reflects the communication overhead of transmitting telemetry and prompts, a crucial factor in IoT networks with strict throughput budgets. Accuracy-related measures (precision, recall, F1-score), as reported in Table~\ref{table:ml_performance_comparison}, quantify detection effectiveness, balancing false positives and false negatives in practical deployments. To ensure statistical robustness, we apply one-way ANOVA followed by Tukey’s HSD test to determine whether differences across LLMs and IDS configurations are significant. This controls Type-I error inflation in multi-comparison settings and provides effect size estimates to evaluate practical impact. Evaluation was conducted under four representative attack scenarios( e.g., DoS, DDoS, brute force, and port scanning), which collectively span availability and reconnaissance threats in IoT deployments. Across 62 independent trials, both standalone ML-based IDS and LLM-augmented hybrids were systematically tested. This design enables a quantification of detection accuracy, interpretability, and operational cost, establishing a strong foundation for cross-attack analysis.

\section{Related Work}
\label{relatedwork}
Recent advances in ML-based IDS have significantly enhanced security capabilities within IoT systems. Specifically, researchers have explored both lightweight detection models, appropriate for resource-limited devices, and more robust inference systems to effectively manage increasingly sophisticated threats. Concurrently, LLMs have begun to demonstrate their potential for enhancing the reasoning and interpretability of alerts generated by IDS. Therefore, this section provides a thorough review of relevant research across ML-based IDS techniques and LLM-enhanced cybersecurity.\\
Otoum et al. \cite{otoum2025llm} present an LLM-driven threat detection and prevention framework specifically designed for IoT. The proposed system integrates lightweight BERT variants, namely TinyBERT, BERT-Mini, and BERT-Small, fine-tuned on domain-specific datasets, including IoT-23 and TON-IoT. Detection outputs are further linked to a rule-based decision engine, enabling real-time mitigation actions. The entire architecture is containerized and deployed in a simulated IoT using Docker, allowing for detailed evaluation under constrained computational conditions. Experimental results indicate that BERT-Small offers the best trade-off between accuracy and efficiency, achieving 99.75\%, detection accuracy with low latency and minimal energy consumption, thereby demonstrating the practical viability of compact LLMs for edge-based intrusion prevention.\\
Diaf et al. \cite{diaf2024beyond} propose a novel intrusion prediction framework that proactively identifies cyber threats in IoT networks by leveraging a combination of fine-tuned LLMs and LSTM. Their architecture integrates GPT-2 for next-packet prediction and BERT for evaluating the coherence of predicted packet sequences, forming a feedback loop that enhances predictive robustness. A final LSTM layer is employed to classify these sequences as either benign or malicious. The framework is trained and tested using the CICIoT2023 dataset, achieving an overall IDS accuracy of 98\%, and demonstrating strong generalization to multistage and unseen attacks.\\
Ferrag et al. \cite{zong2025integrating} introduce security BERT, a privacy-preserving and efficient IDS model tailored for IoT. The framework leverages a modified BERT architecture with a novel BBPE tokenization strategy and employs a BERT-based encoder-decoder structure to analyze network traffic. It is evaluated on the IoTID20 and Edge-IIoT datasets, demonstrating robust detection across multiple types of attacks. SecurityBERT achieves a classification accuracy of 98.2\%, while maintaining extremely low inference time (0.67 ms) and a compact model size (14.3 MB), confirming its suitability for deployment at the edge.\\
Abdelhamid et al. \cite{ferrag2024revolutionizing} present an attention-driven transfer learning framework for IDS in IoT by transforming tabular data into image representations. Their method applies a CNN pre-trained on ImageNet, enhanced with Convolutional Block Attention Modules (CBAM), to extract deep spatial features from visualized network data. The final classification is performed through an ensemble of CNN, RF, and XGBoost models. Evaluated on the CICIoT2023 dataset, the proposed system achieves impressive detection accuracy (up to 99.98\%) while maintaining low false positive rates across various attack types. \\
Adjewa et al. \cite{adjewa2024efficient} propose a federated anomaly-based IDS for 5G-enabled IoT networks using an optimized BERT model tailored for edge deployment. Their architecture reduces the base BERT model to four encoder layers, incorporating linear quantization to compress the model without a significant loss in performance. The system is evaluated using both centralized and federated setups with the Edge-IIoTset dataset. In centralized learning, the model achieves 97.79\% accuracy, while in federated configurations, it reaches up to 97.12\% with IID data and 96.66\% with non-IID data using ten clients. The model also performs effectively on constrained hardware, e.g, Raspberry Pi, with an inference time of 0.45 seconds, underscoring its practical viability for decentralized and privacy-preserving IoT security.\\
Alshamrani et al. \cite{abdelhamid2024attention} propose a two-phase ensemble IDS designed to balance high detection accuracy with computational efficiency for deployment in resource-constrained IoT. The framework employs an Extra Trees classifier and a Deep Neural Network (DNN) for initial binary classification, followed by the RF model to identify specific attack categories. Experiments conducted on the TON-IoT dataset demonstrate that the system achieves competitive accuracy across multiple attacks (e.g., DoS, DDoS), while maintaining a lightweight computational footprint.\\
Aldaej et al.\cite{aldaej2024ensemble} propose a two-stage ensemble IDS framework tailored for IoT-edge platforms, addressing the limitations imposed by resource-constrained devices. In the first stage, an Extra Tree (E-Tree) is used to detect whether incoming IoT traffic is benign and anomalous. In the second stage, an ensemble model comprising E-Tree, DNN, and RF refines the classification by identifying the specific nature of the IDS. The system is evaluated on multiple datasets, including Bot-IoT, CICIDS2018, NSL-KDD, and IoTID20, demonstrating superior performance compared to existing ML techniques.\\

The literature synthesis demonstrates that recent research has achieved notable progress in ML-based IDS, transformer-driven feature extraction, and energy-efficient computing at the edge. Despite these advances, most approaches remain fragmented, lacking integration between detection models and semantic reasoning tools. Few studies have explored the practical combination of lightweight IDS with external LLMs in a deployable architecture. Moreover, the impact of LLMs on enhancing IDS interpretability and decision-making in constrained edge gateways remains largely unexamined.  To address this gap, we benchmark six ML-based IDS models under various cyberattacks at the edge. Their alerts are augmented using three external LLMs (GPT-4-turbo, DeepSeek-V2, and LLaMA~3.5) via real-time API calls. We also evaluate key real-world deployment metrics, such as CPU usage, bandwidth, memory usage, latency, and energy consumption, at the edge. Additionally, we assess the quality of LLM-generated semantic reasoning to determine its practical value for situational awareness.
\section{Proposed Method}
\label{sec:proposed_method}
This section presents the design and methodology of the proposed semantic ML-based IDS deployed at the edge. The approach integrates ML-based IDS with semantically enhanced reasoning via LLMs, forming a hybrid detection pipeline that is both interpretable and resource-efficient. Furthermore, each component is formalized mathematically, followed by detailed explanations that emphasize system performance, security implications, and the practical feasibility of deployment.  
\subsection{Traffic Monitoring and Feature Extraction}
Incoming edge traffic at time $t$ is represented as a structured feature vector:
\begin{equation}
\mathbf{x}_t = [f_1(p_t), f_2(p_t), \dots, f_n(p_t)]^\top, \quad \mathbf{x}_t \in \mathbb{R}^n
\end{equation}
where $f_i$ are feature extraction functions applied to packets $p_t$. Each $f_i$ represents: statistical descriptors (mean packet size, variance), temporal metrics (flow duration, inter-arrival time), and entropy-based measures (protocol diversity, TCP flag entropy). The extracted vector is normalized as:
\begin{equation}
\tilde{\mathbf{x}}_t = \frac{\mathbf{x}_t - \mu}{\sigma}, 
\quad \mu = \frac{1}{T}\sum_{t=1}^T \mathbf{x}_t, 
\quad \sigma = \sqrt{\frac{1}{T}\sum_{t=1}^T (\mathbf{x}_t - \mu)^2}
\end{equation}
where $\mu$ and $\sigma$ are the empirical mean and standard deviation vectors estimated over a baseline benign dataset of size $T$.\\
This normalization ensures strict comparability across heterogeneous edge gateways, regardless of hardware baseline and traffic scale. From a security perspective, normalization protects against adversarial evasion strategies such as \emph{scaling attacks}, in which malicious traffic is deliberately embedded within extreme magnitudes to mislead classifiers. Furthermore, by transforming all features into a standardized space, the IDS prioritizes \emph{relative deviations} (e.g., sudden connection surges or latency spikes) over absolute values, making it significantly harder for attackers to camouflage anomalies inside large flows.  Formally, the mapping is expressed as:
\begin{equation}
\mathcal{N} : \mathbb{R}^n \to \mathbb{R}^n, \quad 
\mathcal{N}(\mathbf{x}_t) = \tilde{\mathbf{x}}_t
\end{equation}
where $\mathcal{N}$ is the normalization operator. This projects $\mathbf{x}_t$ into a standardized feature domain where classification boundaries are invariant to absolute scale.  
Entropy-based features are included as:
\begin{equation}
H(f) = - \sum_{i=1}^{k} p_i \log p_i,
\end{equation}
where $p_i$ denotes the empirical probability of observing protocol or flag category $i$. These measures capture protocol diversity and TCP flag irregularities, improving resilience against polymorphic and obfuscation-based attacks.\\
A major benefit of this formulation is device-agnostic invariance. For example, let $x_{\text{CPU}}^{(s)}$ and $x_{\text{CPU}}^{(g)}$ denote CPU usage for a sensor node and a gateway server, respectively. Although their absolute baselines differ ($\mu^{(s)} \neq \mu^{(g)}$), normalization ensures that an 80\% CPU spike is mapped to the same standardized representation:
\begin{equation}
\tilde{x}_{\text{CPU}}^{(s)} \approx \tilde{x}_{\text{CPU}}^{(g)}.
\end{equation}
This alignment enables a unified IDS policy across heterogeneous edge gateways. From a computational standpoint, the pipeline is bounded by:
\begin{equation}
\mathcal{C}(\mathcal{N}) = \mathcal{O}(n),
\end{equation}
Ensuring predictable runtime even for high-throughput gateways. Linear complexity provides resilience against adversarial attempts at computational flooding: even if attackers inject large volumes of traffic, the system maintains real-time processing without bottlenecks. Additionally, this stage establishes the semantic bridge to the LLM. The normalized feature vectors $\tilde{\mathbf{x}}_t$ form the \emph{structured input layer} for telemetry encoding and prompt construction:
\begin{equation}
\pi_t = \texttt{ENCODE}(\tilde{\mathbf{x}}_t, \mathcal{R}(y_t)),
\end{equation}
Where $\mathcal{R}(y_t)$ maps preliminary ML-based IDS labels into semantic descriptors (see Section~\ref{sec:prompt_construction}). Instead of raw packet traces, the LLM receives compact, security-aware representations that are invariant to device heterogeneity and adversarial scaling, ensuring semantically stable reasoning at the edge.

\subsection{Anomaly Detection using ML-based IDS}
Each normalized vector $\tilde{\mathbf{x}}_t$ is evaluated by a trained classifier $f_\theta$:
\begin{equation}
y_t = f_\theta(\tilde{\mathbf{x}}_t), \quad s_t \in [0,1]
\end{equation}
Where $y_t$ is the predicted traffic class label and $s_t$ is the calibrated anomaly score. The anomaly score is defined as:
\begin{equation}
s_t = 1 - P(y_t = \text{benign} \mid \tilde{\mathbf{x}}_t, \theta),
\end{equation}
Representing the probability mass assigned to non-benign hypotheses. This formulation can be interpreted as a generalized likelihood-ratio test:
\begin{equation}
\Lambda(\tilde{\mathbf{x}}_t) = \frac{P(\tilde{\mathbf{x}}_t \mid H_1)}{P(\tilde{\mathbf{x}}_t \mid H_0)}, 
\quad \text{trigger alert if } \Lambda(\tilde{\mathbf{x}}_t) \geq \eta
\end{equation}
Where $H_0$ and $H_1$ correspond to benign and attack hypotheses, respectively.  For classifiers with probabilistic outputs (e.g., logistic regression, random forest, neural networks with softmax layers), the anomaly score is reformulated as:
\begin{equation}
s_t = \max_{y \in \mathbb{Y} \setminus \{\text{benign}\}} P(y_t = y \mid \tilde{\mathbf{x}}_t, \theta),
\end{equation}
This reflects the most probable malicious hypothesis against the benign baseline. In practice, this acts as a \emph{security margin}: higher $s_t$ indicates more substantial evidence of malicious behavior.  For margin-based classifiers, anomaly scores are defined as:
\begin{equation}
s_t = \sigma(\mathbf{w}^\top \tilde{\mathbf{x}}_t + b),
\end{equation}
where $\mathbf{w}$ is the separating hyperplane, $b$ is the bias, and $\sigma(\cdot)$ is a sigmoid calibration function. This ensures anomaly scores are smoothly mapped into $[0,1]$, making them compatible with threshold-based decision policies.  From a security perspective, this stage represents the system’s first line of defense. The generalized likelihood-ratio test formulation enforces bounded false-alarm rates ($P_{FA} \leq \alpha$ for threshold $\eta$), ensuring that benign background noise does not overwhelm downstream reasoning layers. This statistical grounding is critical in adversarial environments where attackers may attempt to inject \emph{camouflage traffic} to mimic benign distributions. From a computational perspective, anomaly detection operates with complexity:
\begin{equation}
\mathcal{O}(n \cdot d),
\end{equation}
Where $n$ is the number of input features and $d$ represents model depth (tree depth for decision trees and number of hidden layers for neural models). This guarantees scalability for edge-deployed ML-based IDS under high-frequency feature extraction. In addition, this design inherently supports adversarial robustness. Since $s_t$ is derived from probabilistic margins rather than raw thresholds, adversaries manipulate multiple correlated features simultaneously to evade detection. This raises the cost of evasion and enhances resilience against poisoning and adversarial bypass strategies, ensuring reliable anomaly detection at the edge.

\subsection{Telemetry Collection and Normalization}
Once the anomaly score $s_t$ exceeds the predefined alert threshold, the system transitions into telemetry collection. At this stage, a compact system-state vector is constructed:
\begin{equation}
\mathbf{m}_t = [c_t, m_t, l_t, e_t, s_t]^\top
\end{equation}
where $c_t$ represents CPU usage, $m_t$ denotes memory usage, $l_t$ captures latency, $e_t$ indicates energy consumption, and $s_t$ reflects the anomaly score inherited from the detection stage.
To ensure device independence, the raw telemetry vector is normalized against a baseline capacity vector $\bar{\mathbf{m}}$:
\begin{equation}
\tilde{\mathbf{m}}_t = \mathbf{m}_t \oslash \bar{\mathbf{m}}, 
\quad \bar{\mathbf{m}} = [100,\,2048,\,50,\,300,\,1]^\top
\end{equation}
This guarantees consistency across heterogeneous platforms. For example, a Raspberry Pi reporting 80\% CPU usage maps to a normalized value of $0.80$. In effect, normalization eliminates bias arising from absolute hardware differences, thereby enabling fair semantic reasoning across devices.\\
Formally, each component of the telemetry vector is scaled as:
\begin{equation}
\tilde{m}_{t,i} = \frac{m_{t,i}}{\bar{m}_i}, \quad i \in \{1,\dots,5\}
\end{equation}
Ensuring that all dimensions are bounded within $[0,1]$. This introduces scale invariance, which shifts detection logic from raw resource magnitudes toward relative deviations. Consequently, reasoning becomes device-agnostic, focusing on anomalous utilization rather than absolute specifications. From a statistical perspective, the normalized telemetry vector can be modeled as a point in a five-dimensional unit hypercube:
\begin{equation}
\tilde{\mathbf{m}}_t \in [0,1]^5
\end{equation}
This compact embedding enables efficient similarity operations between telemetry snapshots across devices, allowing for comparative anomaly tracking at scale. From a security perspective, normalization thwarts evasion attempts that exploit device heterogeneity. Without normalization, an attacker might saturate a weaker IoT sensor while remaining undetected on a stronger edge server. Moreover, by mapping all observations into the same normalized scale, such tactics are neutralized, since anomaly detection now depends only on proportional deviations.  From a computational perspective, telemetry collection is lightweight, with complexity $\mathcal{O}(k)$ where $k=5$ metrics are monitored. The normalization itself requires only vector division, incurring insignificant cost and latency well below a millisecond, even across thousands of devices.  Additionally, the normalized telemetry vector $\tilde{\mathbf{m}}_t$ serves as the \emph{core evidence} for subsequent LLM-based reasoning. It provides a mathematically bounded, semantically interpretable summary of system health that is fused with contextual attack descriptors during prompt construction. In this way, telemetry normalization not only harmonizes cross-device monitoring but also establishes the foundation for robust semantic reasoning at the edge.

\subsection{Prompt Construction}
\label{sec:prompt_construction}
Following telemetry normalization, the system transitions into prompt construction, where the evidence is transformed into a semantically interpretable input for the LLM. The encoded prompt is defined as:
\begin{equation}
\pi_t = \texttt{ENCODE}(\tilde{\mathbf{m}}_t, \mathcal{R}(y_t))
\end{equation}
Here, $\tilde{\mathbf{m}}_t$ represents the normalized telemetry vector, while $\mathcal{R}(y_t)$ maps each predicted attack class to a textual description drawn from a fixed, domain-specific knowledge base. This ensures that quantitative anomalies are seamlessly aligned with qualitative semantic context.\\
To maintain interpretability and efficiency, the final prompt is structured into three ordered blocks: (i) a snapshot of telemetry metrics, (ii) the contextualized description of the predicted attack, and (iii) explicit reasoning instructions guiding the LLM toward actionable diagnostics. Formally, the prompt is bounded in size by:
\begin{equation}
|\pi_t| \leq B_{\max}, \quad B_{\max} = 1.2\,\text{kB}
\end{equation}
This upper bound ensures that prompts remain lightweight enough for low-bandwidth communication channels. From an optimization perspective, prompt construction can be expressed as a constrained concatenation problem:
\begin{equation}
\pi_t = \arg\min_{\pi} \Big( \mathcal{L}_{\text{semantic}}(\pi) \Big), 
\quad \text{s.t. } |\pi| \leq B_{\max}
\end{equation}
Where $\mathcal{L}_{\text{semantic}}(\pi)$ penalizes information loss due to truncation, ensuring that critical attack context is always preserved under bandwidth limitations.\\
From a security perspective, prompt construction serves as a safeguard against ambiguous or adversarial signals. For example, a raw metric such as ``CPU = 0.92'' is semantically tied to its associated threat domain (e.g., ``brute force with repeated failed logins'') through $\mathcal{R}(y_t)$. This mapping prevents misinterpretation and constrains the LLM to reason strictly within the appropriate attack context. Furthermore, since $\pi_t$ is both compact and normalized, it reduces the attack surface for prompt injection or adversarial tampering during transmission.\\
Additionally, prompts can be treated as structured embeddings in a semantic vector space, enabling retrieval-based reasoning. Specifically, similarity between two prompts is defined as:
\begin{equation}
\text{sim}(\pi_t, \pi_{t'}) = \frac{\langle \pi_t, \pi_{t'} \rangle}{\|\pi_t\|\|\pi_{t'}\|}
\end{equation}
This formulation allows the system to retrieve historical prompts most similar to the current one, thereby supporting few-shot reasoning without maintaining heavy persistent databases. As a result, the IDS inherits a form of lightweight semantic memory, improving adaptability to recurring attack patterns. From a computational standpoint, prompt construction is efficient: it requires $\mathcal{O}(k)$ operations for telemetry encoding, where $k$ is the number of monitored metrics, and $\mathcal{O}(1)$ for context mapping. This makes its overhead negligible compared to IDS inference, ensuring that the pipeline can operate seamlessly on constrained edge hardware. Besides, prompt construction acts as the semantic bridge between quantitative system telemetry and qualitative reasoning. It compresses heterogeneous signals into a bandwidth-constrained yet context-rich representation, enabling the LLM to generate interpretable, attack-specific insights that extend far beyond raw anomaly scores.
\subsection{Remote Reasoning via LLM API}
Once the prompt is constructed, it is transmitted to the remote LLM for semantic reasoning. The inference process is defined as:
\begin{equation}
\mathcal{L}_\star(\pi_t) \mapsto \langle \hat{y}_\star, \gamma_\star, \mu_\star, \rho_\star \rangle
\end{equation}
Where $\hat{y}_\star$ denotes the refined attack classification, $\gamma_\star \in [0,1]$ is the calibrated semantic confidence, $\mu_\star$ encodes mitigation recommendations, and $\rho_\star \in \{\text{Normal}, \text{Warning}, \text{Critical}\}$ specifies the severity level.\\
In this stage, the raw anomaly score $s_t$ is elevated into \emph{actionable cybersecurity intelligence}. For instance, instead of a low-level alert such as ``anomaly score = 0.95,'' the LLM output a structured message: \emph{``Brute force attack detected, confidence = 92\%, severity = Critical, mitigation = lockout + MFA.''} This semantic enrichment bridges the gap between statistical detection and operator-ready defense policies.  To preserve trustworthiness, each prompt is signed before transmission:
\begin{equation}
h(\pi_t) = \texttt{HASH}(\pi_t)
\end{equation}
And acceptance is conditioned on integrity verification:
\begin{equation}
h(\pi_t) = h(\pi_t^{\text{received}})
\end{equation}
Otherwise, the request is discarded. This mechanism ensures resilience against prompt tampering, replay attacks, and attempts at adversarial injection.\\
From a probabilistic viewpoint, the confidence $\gamma_\star$ can be interpreted as:
\begin{equation}
\gamma_\star = P(\hat{y}_\star \mid \pi_t, \Theta_{\text{LLM}})
\end{equation}
where $\Theta_{\text{LLM}}$ are the model parameters. This posterior formulation provides a principled measure of reliability, allowing operators to enforce strict acceptance thresholds (e.g., $\gamma_\star \geq \gamma_{\min}$). Consequently, uncertain inferences are either escalated for human review or filtered out to avoid false mitigations. \\
From a security perspective, remote reasoning adds multi-step causal inference that traditional IDS lacks. By synthesizing telemetry, context, and prior attack knowledge, the LLM generates adaptive and context-aware defense strategies. At the same time, new risks emerge: adversaries may attempt adversarial prompt engineering or resource-exhaustion attacks. To mitigate this, integrity hashing, semantic thresholds, and rate-limiting policies are embedded into the pipeline, ensuring that only validated and high-confidence requests consume LLM resources.  From a computational perspective, LLM calls represent the dominant contributor to both latency and energy consumption. To balance scalability with security, these calls are activated selectively: only when the anomaly score $s_t$ surpasses the alert threshold $\tau_{\text{alert}}$. This conditional invocation ensures that semantic reasoning is reserved for suspicious traffic, thereby maintaining efficiency across hundreds of concurrent edge gateways. Additionally, this stage transforms raw statistical detections into semantically grounded, context-aware security responses. It secures the integrity of transmitted prompts, quantifies reliability via calibrated confidence, and integrates LLM-driven reasoning into a resource-aware IDS pipeline suitable for the edge gateway.
\subsection{Inference Pipeline}
The whole system pipeline is summarized in Algorithm~\ref{alg:semantic_ids}. This algorithm provides a structured workflow that integrates traditional ML-based IDS with semantic reasoning through LLMs. At the start, incoming traffic is transformed into normalized feature vectors to ensure consistency across heterogeneous edge gateways and to prevent adversaries from exploiting statistical biases. Once features are prepared, the ML-based IDS performs local classification and assigns an anomaly score $s_t$, which serves as a primary defense filter by preventing benign traffic from triggering unnecessary semantic analysis. When suspicious activity is detected, runtime telemetry is collected and normalized, ensuring that CPU, memory, latency, energy, and anomaly values remain comparable across devices of different capacities.
This guarantees fairness in interpretation and prevents attackers from exploiting hardware-specific weaknesses. The prompt construction stage then embeds both the telemetry and contextual information about the predicted attack type, producing a semantically rich query for the LLM. Importantly, before transmission, the system applies a hashing function $h(\pi_t)$ to the prompt, ensuring integrity verification and safeguarding against tampering or adversarial prompt injection during communication. Once the LLM returns its enriched response, the system applies a validation mechanism that discards outputs with low semantic confidence values $\gamma_\star < \gamma_{\min}$. This security check ensures that unreliable or adversarially manipulated responses are not automatically acted upon. The final findings, including refined attack labels, severity ratings, and recommended mitigations, are logged and forwarded both to human analysts and automated defense actuators. Additionally, Algorithm~\ref{alg:semantic_ids} is not only a pipeline for semantic enrichment but also a layered defense mechanism. It combines anomaly filtering, telemetry normalization, cryptographic hashing, and semantic validation, thereby enhancing robustness against adversarial noise, data poisoning, and injection attacks. 
\begin{algorithm}[H]
\caption{LLM-Integrated ML-based IDS Pipeline}
\label{alg:semantic_ids}
\begin{algorithmic}[1]
\Require Feature stream $\{\mathbf{x}_t\}$, trained IDS $f_\theta$, threshold $\tau_{\text{alert}}$, context map $\mathcal{R}(\cdot)$, LLM API $\mathcal{L}_\star$
\For{each time step $t$}
    \State Extract features $\mathbf{x}_t \gets \texttt{Extract}(p_t)$
    \State Normalize $\tilde{\mathbf{x}}_t \gets (\mathbf{x}_t - \mu)/\sigma$
    \State Predict class $y_t \gets f_\theta(\tilde{\mathbf{x}}_t)$
    \State Compute anomaly score $s_t$
    \If{$s_t \geq \tau_{\text{alert}}$}
        \State Collect telemetry $\mathbf{m}_t \gets [c_t,m_t,l_t,e_t,s_t]$
        \State Normalize $\tilde{\mathbf{m}}_t \gets \mathbf{m}_t \oslash \bar{\mathbf{m}}$
        \State Retrieve context $ctx_t \gets \mathcal{R}(y_t)$
        \State Construct $\pi_t \gets \texttt{ENCODE}(\tilde{\mathbf{m}}_t, ctx_t)$
        \State Sign $h(\pi_t) \gets \texttt{HASH}(\pi_t)$
        \State Query LLM $\langle \hat{y}_\star,\gamma_\star,\mu_\star,\rho_\star \rangle \gets \mathcal{L}_\star(\pi_t)$
        \State Validate: reject if $\gamma_\star < \gamma_{\min}$
        \State Log outputs, trigger mitigations
    \Else
        \State Continue benign logging
    \EndIf
\EndFor
\end{algorithmic}
\end{algorithm}
\subsection{Latency and Energy Modeling}
The end-to-end detection delay is formalized as:
\begin{equation}
T_t = T_{\text{IDS}} + T_{\text{tx}} + T_{\text{LLM}}
\end{equation}
Where $T_{\text{IDS}}$ denotes the local ML-based inference time (typically measured in milliseconds), $T_{\text{tx}}$ represents the communication delay between the edge gateway and the remote reasoning service, and $T_{\text{LLM}}$ accounts for the dominant latency introduced by semantic reasoning This decomposition enables precise identification of bottlenecks within the pipeline. For instance, lightweight classifiers reduce $T_{\text{IDS}}$, whereas limited-bandwidth edge networks often dominate $T_{\text{tx}}$. In contrast, complex reasoning over large prompts primarily increases $T_{\text{LLM}}$. From a security perspective, bounding $T_t$ is essential: adversaries exploit delayed responses to saturate resources before mitigations are applied, thereby amplifying attack impact. Consequently, latency modeling serves not only as a performance metric but also as a \emph{defensive safeguard} against timing-based adversarial strategies. The energy consumption per detection cycle is expressed as:
\begin{equation}
E_t = \sum_{i=1}^{N} P_i \cdot \Delta\tau, 
\quad \Delta\tau = 10\,\text{ms}, 
\quad N = \frac{T_t}{\Delta\tau}
\end{equation}
Where $P_i$ denotes instantaneous power at sub-interval $i$, this discrete-time integration offers fine-grained profiling of device energy usage during the inference and reasoning cycle. Crucially, the model highlights that \emph{latency and energy are tightly coupled}; higher $T_{\text{LLM}}$ simultaneously increases $T_t$ and amplifies $E_t$. For edge devices with constrained batteries, this trade-off determines long-term sustainability. From a defensive perspective, deviations in $E_t$ can signal energy-drain attacks, where adversaries deliberately trigger costly LLM invocations to exhaust device resources. By continuously monitoring $E_t$ relative to expected baselines, the IDS gains an additional detection dimension that complements anomaly scoring. This ensures that resilience is not limited to traffic-level anomalies but extends to system-level sustainability.
\subsection{Operational Constraints}
To ensure resilience under adversarial and resource-limited conditions, the IDS is governed by strict operational constraints:
\begin{align}
T_t &\leq T_{\max} = 1.5\,\text{s} \\
E_t &\leq E_{\text{budget}} = 100\,\text{J} \\
\gamma_\star &\geq \gamma_{\min} = 0.60
\end{align}
Constraint (1) enforces near-real-time responsiveness by bounding the end-to-end detection latency. This ensures that attack mitigation actions are applied quickly enough to prevent escalation, even under high traffic loads or adversarial attempts to delay system responses. Constraint (2) restricts the per-cycle energy consumption of the IDS, thereby protecting battery-powered and resource-constrained edge gateways against energy-drain attacks. By maintaining $E_t \leq 100\,\text{J}$, the system guarantees sustainable operation without sacrificing detection capability.  Constraint (3) introduces a semantic confidence floor on LLM-based reasoning. Any refined prediction with $\gamma_\star < \gamma_{\min}$ is discarded, ensuring that low-certainty outputs do not trigger misleading and unsafe mitigation actions. This guards against adversarial prompt manipulation that attempts to generate ambiguous or low-confidence outputs. Moreover, these bounds define a constrained optimization problem:
\begin{equation}
\max \; \text{Detection Accuracy}, 
\quad \text{s.t. } (1) \wedge (2) \wedge (3)
\end{equation}
Where the IDS strives to maximize detection performance while simultaneously respecting latency, energy, and reliability budgets, from a system-level perspective, these constraints operate as runtime defenses. They prevent adversaries from bypassing security by overwhelming the IDS with timing or resource-based attacks and guarantee stable operation across heterogeneous edge gateways. By embedding these safeguards directly into the optimization framework, the IDS balances \emph{accuracy, efficiency, and robustness}, ensuring practicality for deployment at the edge.
\subsection{Evaluation Framework}
We evaluate six ML-based IDS configurations:
\begin{equation}
f_\theta \in \{\text{DT}, \text{KNN}, \text{RF}, \text{CNN}, \text{LSTM}, \text{CNN+LSTM}\}
\end{equation}
Across five cyber attack classes:
\begin{equation}
\mathbb{Y}' = \{\text{DoS}, \text{DDoS}, \text{brute force}, \text{port scanning}\}
\end{equation}
Evaluation is based on four dimensions:
\begin{itemize}
    \item \textbf{Accuracy:} Standard classification metrics (precision, recall, F1 score), measuring raw ML-based IDS effectiveness.
    \item \textbf{Semantic gain:} Improvement $\Delta F_1$ after LLM reasoning, quantifying interpretability and contextual awareness.
    \item \textbf{Runtime:} Average latency $T_t$ and energy $E_t$, validating efficiency under attack stress.
    \item \textbf{Interpretability:} Human expert rating $\iota \in [1,5]$, assessing how actionable and understandable the enriched outputs are.
\end{itemize}
This multi-metric framework goes beyond accuracy to capture interpretability, efficiency, and resilience. Moreover, by analyzing $\Delta F_1$, it quantifies the semantic contribution of LLMs in reducing false positives and improving operator trust. Runtime and energy measurements verify that constraints (1)–(3) are met even under resource-intensive attacks. Interpretability scores confirm that the enriched outputs bridge the gap between raw anomaly signals and practical security decision-making. Collectively, this framework validates the feasibility of deploying the hybrid IDS at the IoT edge, ensuring both technical and security robustness.
\section{Experimental design}
\label{Study design}
This section describes our methodology for evaluating the impact of integrating ML-based IDSs with LLMs on the edge gateway. We first present our Research Questions (RQs), followed by an explanation of the study design, which includes the deployment environment, traffic scenarios, and evaluation metrics used to analyze detection performance.
\subsection{Research questions(RQs)}
\label{Research questions(RQs)}
 Our research aims to address the following RQs: 
\begin{itemize}
    \item \textbf{RQ1: How does the integration of LLMs impact the detection accuracy of ML-based IDSs under different cyberattacks?}\\
    This question examines whether LLM-enhanced ML-based IDS can achieve measurable improvements in classification accuracy across various types of cyberattacks.
    \item \textbf{RQ2: To what extent can external LLMs enhance the semantic reasoning and interpretability of ML-based IDS outputs without compromising real-time detection performance?}\\
    This question examines how effectively LLMs improve the clarity and contextual relevance of ML-based IDS alerts while maintaining operational responsiveness for real-world applications.
    \item \textbf{RQ3: What is the impact of LLM-assisted ML-based IDS on system-level performance metrics, e.g, CPU usage, energy consumption, and latency when deployed at the edge?}\\
    This question assesses the resource overhead introduced by incorporating LLMs and evaluates whether such integration respects the real-time constraints of low-power platforms( e.g., Raspberry Pi).
\end{itemize}
\subsection{Dataset}
\label{Dataset}
We used the CICIDS2017 dataset~\cite{stiawan2020cicids}, a widely accepted benchmark that simulates realistic network traffic, including both benign behavior and diverse cyberattacks. Its comprehensive coverage makes it ideal for evaluating the performance of ML-based IDS in IoT systems. The distribution of attacks and benign entries is detailed in Table~\ref{tab: dataset_distribution}.
\begin{table}[h]
\centering
\caption{Distribution of labeled IoT attacks in the dataset}
\begin{tabular}{|l|r|}
\hline
\textbf{IoT Attack Labels}        & \textbf{No of Labeled Entries} \\ \hline
BENIGN                      & 2,271,320 \\ \hline
DoS Hulk                    & 230,124   \\ \hline
Port Scan                   & 158,804   \\ \hline
DDoS                        & 128,025   \\ \hline
DoS GoldenEye               & 10,293    \\ \hline
FTP-Patator                 & 7,935     \\ \hline
SSH-Patator                 & 5,897     \\ \hline
DoS Slowloris               & 5,796     \\ \hline
DoS Slowhttptest            & 5,499     \\ \hline
Bot                         & 1,956     \\ \hline
Web Attack \& Brute Force   & 1,507     \\ \hline
Web Attack \& XSS           & 652       \\ \hline
Infiltration                & 36        \\ \hline
Web Attack \& SQL Injection & 21        \\ \hline
Heartbleed                  & 11        \\ \hline
\end{tabular}
\label{tab: dataset_distribution}
\end{table}
\subsection{Implementation details}
\label{Experimental design}
\begin{figure*}[!t]
  \centering
  \includegraphics[width=.65\textwidth]{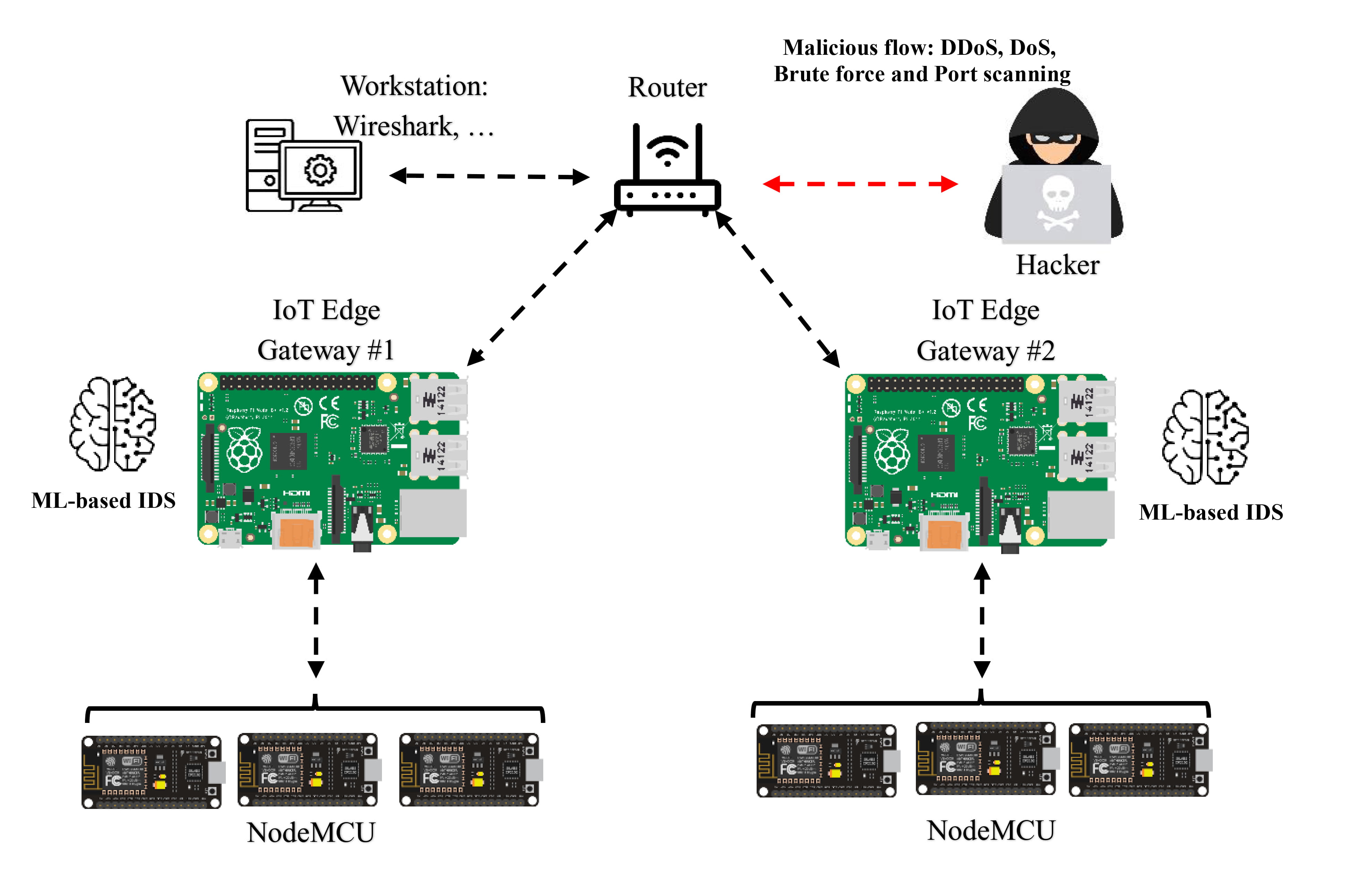}
  \caption{IoT-edge testbed topology.}
  \label{fig:SDN_and_IoT}
\end{figure*}

To assess the practicality of deploying a semantically ML-based IDS at the edge, we designed and implemented an experimental testbed using two Raspberry Pi 4 Model B units, each equipped with 8 GB of RAM and a 1.5 GHz 64-bit quad-core processor. This configuration reflects the constraints of a real-world edge gateway, allowing for a thorough evaluation of resource consumption and operational performance. On these devices, we deployed six widely adopted ML-based IDS for real-time IDS. When an anomaly is identified, the edge device collects a telemetry snapshot comprising CPU and memory usage, latency, energy consumption, and the model’s anomaly score. This information is normalized and transmitted via secure, low-bandwidth API calls to external LLMs to generate interpretable threat explanations and recommended mitigation actions. Furthermore, the system was evaluated using real-time cyber threats by Kali Linux. The overall architecture of the IoT testbed is illustrated in Figure~\ref{fig:SDN_and_IoT}.
\section{Experimental results}
\label{Experimental results and analysis}
To address RQ1 and RQ2, we designed a controlled evaluation framework that tests the effectiveness of our composable prompt architecture. Specifically, four representative cyberattacks were simulated and applied across three LLMs(GPT-4-turbo, DeepSeek-V2, and LLaMA 3.5), ensuring fairness by providing each model with identical telemetry inputs, static RAG-based context, and standardized instruction blocks. In addition, every LLM was evaluated under three distinct reasoning modes (zero-shot, few-shot, and chain-of-thought), enabling a systematic analysis of how prompting strategies influence detection accuracy, interpretability, and resource efficiency.  

\subsection{Brute Force Attack Evaluation}
\label{sec:bf_attack_evaluation}
This section evaluates the detection of brute-force attacks using ML-based IDS integrated with three LLMs (GPT-4-turbo, DeepSeek-V2, and LLaMA3.5) at the edge. The analysis focuses on three main aspects: (i) the reasoning quality and interpretability of each LLM, (ii) the resource consumption patterns under brute-force attack, and (iii) the statistical validity of observed variations across IDS and LLM configurations. Moreover, by presenting a holistic evaluation, we highlight both the strengths and weaknesses of hybrid IDS-LLM frameworks in handling high-frequency brute-force intrusions.  
\subsubsection{Security and Reasoning Quality}
\begin{figure*}[htbp]
    \centering
    \includegraphics[width=0.9\linewidth]{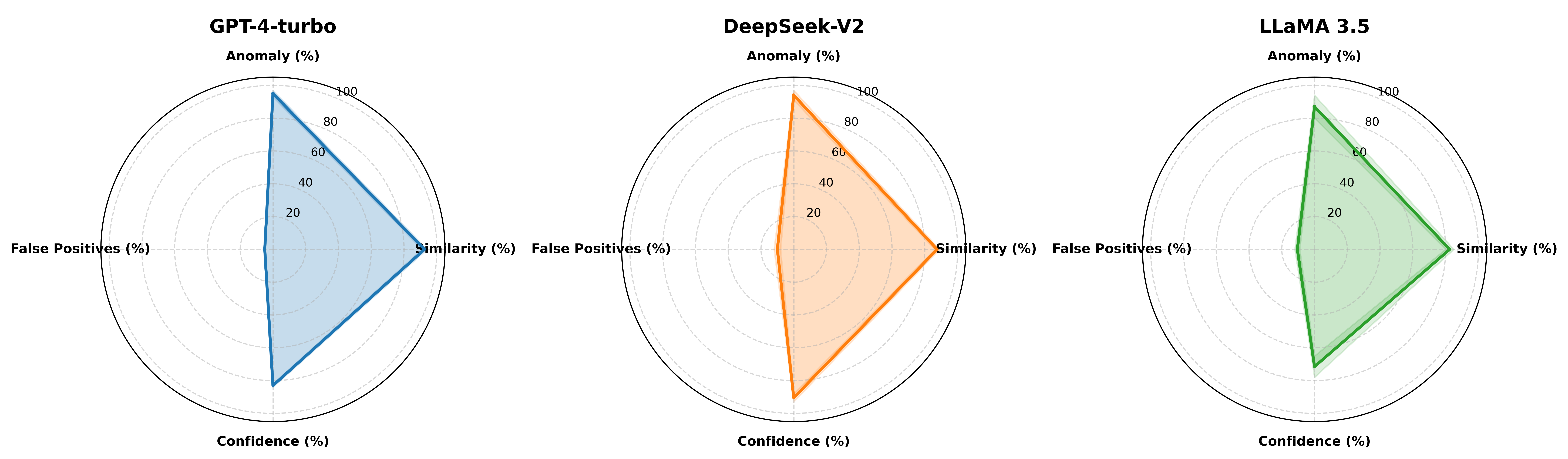}
    \caption{Security reasoning performance under brute-force.}
    \label{fig:bf_reasoning}
\end{figure*}

Brute-force attacks are characterized by repeated authentication failures, which makes reasoning quality and interpretability of IDS responses particularly critical. As illustrated in Figure~\ref{fig:bf_reasoning}, across all IDS models, GPT-4-turbo exhibited the most coherent and actionable reasoning. In few-shot mode, it consistently aligned anomaly scores (93-97\%) and detected connection attempts (300-370) with brute-force behavioral signatures, thereby yielding similarity scores of 91-94\% while maintaining a false positive rate of 5\%. Moreover, its CoT reasoning was both structured and multi-modal, as it integrated anomaly scores, CPU usage (70-82\%), and latency values (9-11 ms) to produce confidence levels of up to 84\%. In addition, GPT-4-turbo generated semantically meaningful mitigation strategies, including multi-factor authentication (MFA), lockout policies, and IP blocking.\\  
By contrast, DeepSeek-V2 also demonstrated substantial numerical precision, with similarity levels between 86-89\% and confidence scores often exceeding 90\%. However, its reasoning remained formulaic, as the system heavily emphasized CPU usage (70-92\%) and anomaly scores (91-97\%), while underweighting secondary telemetry, such as bandwidth fluctuations ($<$95 MB/s). Consequently, this one-dimensional reasoning approach occasionally limited its interpretability, as secondary indicators such as connection diversity were underrepresented. As a result, false positive rates were higher (8-12\%), suggesting that while DeepSeek-V2 is reliable in detecting brute-force attempts, its context awareness and cross-metric integration remain limited compared to GPT-4-turbo.\\  
On the other hand, LLaMA3.5 consistently ranked lowest in reasoning quality. Its outputs often isolated single features such as latency (10-13 ms) and memory usage ($>$1150 MB) without contextual integration. Although structurally compliant with reasoning formats, it nevertheless failed to establish causal relationships across metrics. Consequently, similarity scores remained lower (80-85\%), confidence frequently dropped to 65-78\%, and in some instances, severe brute-force attacks were downgraded to ``warnings.'' This conservative bias. However, reducing false negatives in some scenarios poses serious risks in security-critical scenarios at the edge, where underestimating an attack can directly compromise device integrity and system safety.

\subsubsection{Resource Consumption }
\begin{figure*}[htbp]
    \centering
    \includegraphics[width=0.9\linewidth]{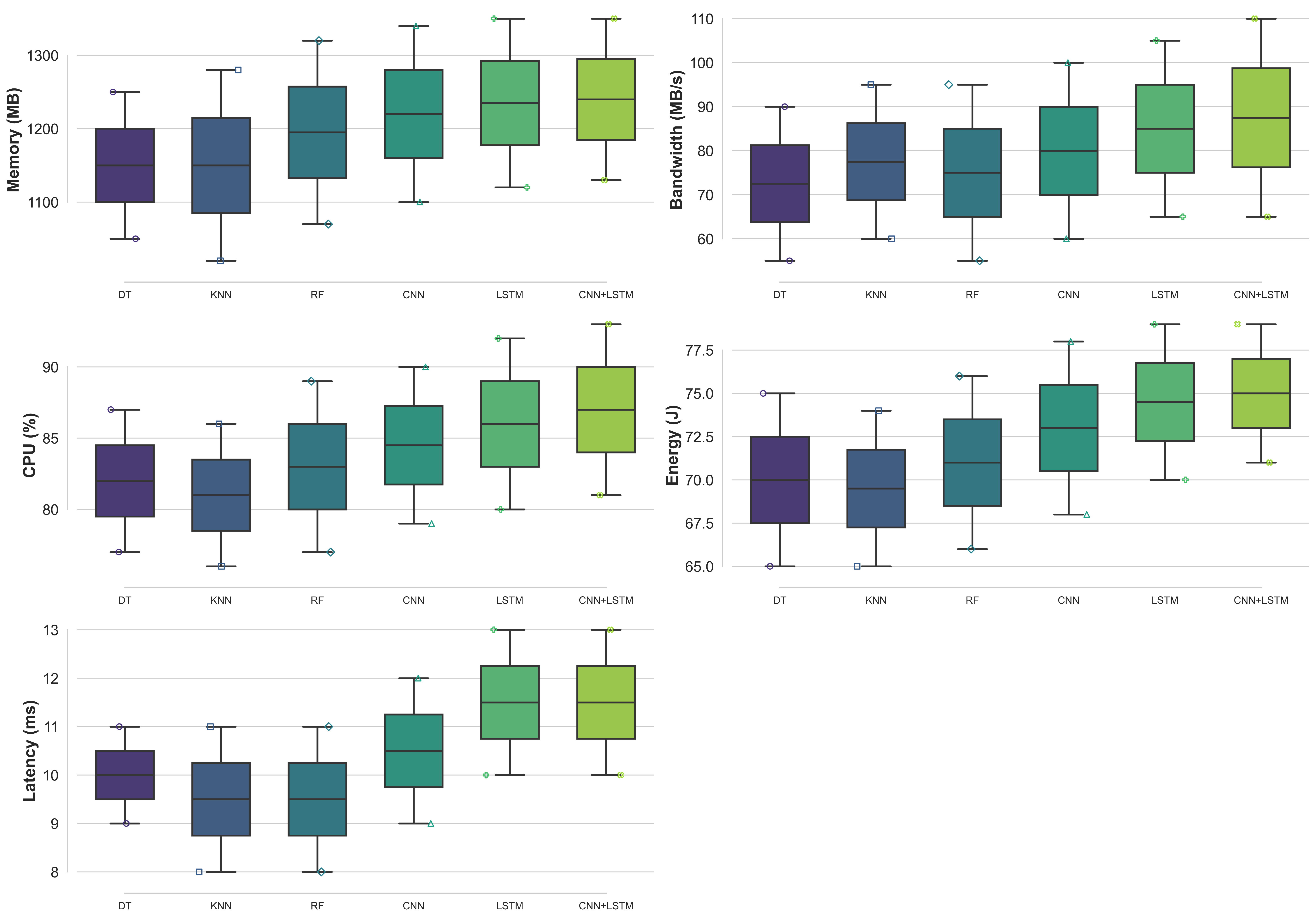}
    \caption{Resource usage under brute-force.}
    \label{fig:bf_resources}
\end{figure*}

Resource utilization during brute-force detection revealed essential insights into the efficiency of ML-based IDS-LLM integration. As shown in Figure~\ref{fig:bf_resources}, classical ML-based IDS models such as DT, KNN, and RF consistently demonstrated resource efficiency, since they required lower memory, CPU, and energy demands. Specifically, these models consumed 1020-1320 MB of memory, maintained CPU usage between 76-89\%, and produced latency values in the range of 8-11 ms, thereby making them suitable for resource-constrained edge gateways.\\  
By contrast, deeper ML-based IDS models (CNN, LSTM, and a hybrid model of CNN and LSTM) imposed significantly higher overhead. Memory usage peaked at 1340 MB, CPU usage ranged from 79-93\%, and energy consumption reached up to 79 J. Moreover, latency also increased (10-13 ms), reflecting the computational complexity of DL architectures. Interestingly, the integration of LLMs did not drastically amplify resource consumption beyond the base ML-based IDS requirements. Instead, the LLM component primarily impacted interpretability and reasoning quality, while imposing insignificant additional overhead on memory, CPU, and bandwidth.\\  
Consequently, the resource trade-off highlights a fundamental deployment challenge; while CNN and LSTM-based IDS improve raw detection accuracy, they also increase operational costs. For edge, this implies that selecting the ML-based IDS backbone should depend not only on accuracy but also on the available device capabilities and energy budgets. Furthermore, the LLM choice (GPT-4-turbo, DeepSeek-V2, or LLaMA3.5) had only a marginal impact on system-level resource footprints.

\subsubsection{Statistical Analysis}
Statistical evaluation using one-way ANOVA and Tukey HSD confirmed that differences in brute-force detection overhead across LLMs were not statistically significant. Specifically, as summarized in Table~\ref{tab:bf_stats}, for all five resource metrics (memory, bandwidth, CPU, energy, and latency), $p$-values exceeded 0.05 and effect sizes remained insignificant ($\eta^2 < 0.003$). Consequently, this indicates that LLM selection does not materially impact resource consumption, even under sustained brute-force scenarios. Moreover, since LLM choice does not introduce significant computational costs, deployment decisions can instead be guided by reasoning quality, interpretability, and security robustness rather than raw performance efficiency. Thus, the emphasis shifts toward evaluating how effectively each LLM translates raw telemetry into semantically actionable insights.
\begin{table*}[!t]
\centering
\caption{Statistical analysis under brute-force.}
\label{tab:bf_stats}
\renewcommand{\arraystretch}{1.2}
\begin{tabular}{|l|c|c|c|c|}
\hline
Metric & ANOVA F & p-value & Effect Size ($\eta^2$) & Tukey Result \\
\hline
Memory     & 0.868 & 0.420 & 0.0016 & No difference \\
Bandwidth  & 0.106 & 0.899 & 0.0002 & No difference \\
CPU        & 1.483 & 0.227 & 0.0027 & No difference \\
Energy     & 0.106 & 0.899 & 0.0002 & No difference \\
Latency    & 1.182 & 0.307 & 0.0022 & No difference \\
\hline
\end{tabular}
\end{table*}

\begin{tcolorbox}[
  colback=green!10!white,
  colframe=green!60!black,
  coltitle=white,
  fonttitle=\bfseries,
  title=Finding,
  enhanced,
   breakable,         
  drop shadow
]
GPT-4-turbo delivered the strongest balance of reasoning depth, interpretability, and actionable mitigation strategies, making it the most security-relevant choice for brute-force detection. DeepSeek-V2 offered reliable anomaly detection with strong numerical accuracy but remained formulaic and less context-aware. LLaMA3.5, while lightweight, provided shallow reasoning and, in some cases, downgraded critical intrusions, creating risks for a mission-critical edge gateway. Since resource utilization did not significantly differ across LLMs, the decisive factor in ML-based IDS-LLM integration lies in the quality of semantic reasoning rather than computational overhead.
\end{tcolorbox}

\subsection{DoS Attack Evaluation}
\label{sec:dos_attack_evaluation}
This section evaluates DoS detection under LLM-integrated ML-based IDS, focusing on three significant aspects: (i) the reasoning quality and interpretability of LLM outputs, (ii) the resource consumption patterns when subjected to DoS, and (iii) the statistical significance of observed differences across models. Since DoS attacks aim to saturate system resources from a single source, both the semantic reasoning ability of LLMs and the efficiency of underlying ML-based IDS backbones play a crucial role in sustaining detection performance under stress.
\subsubsection{Security and Reasoning Quality}
\begin{figure*}[htbp]
    \centering
    \includegraphics[width=0.9\linewidth]{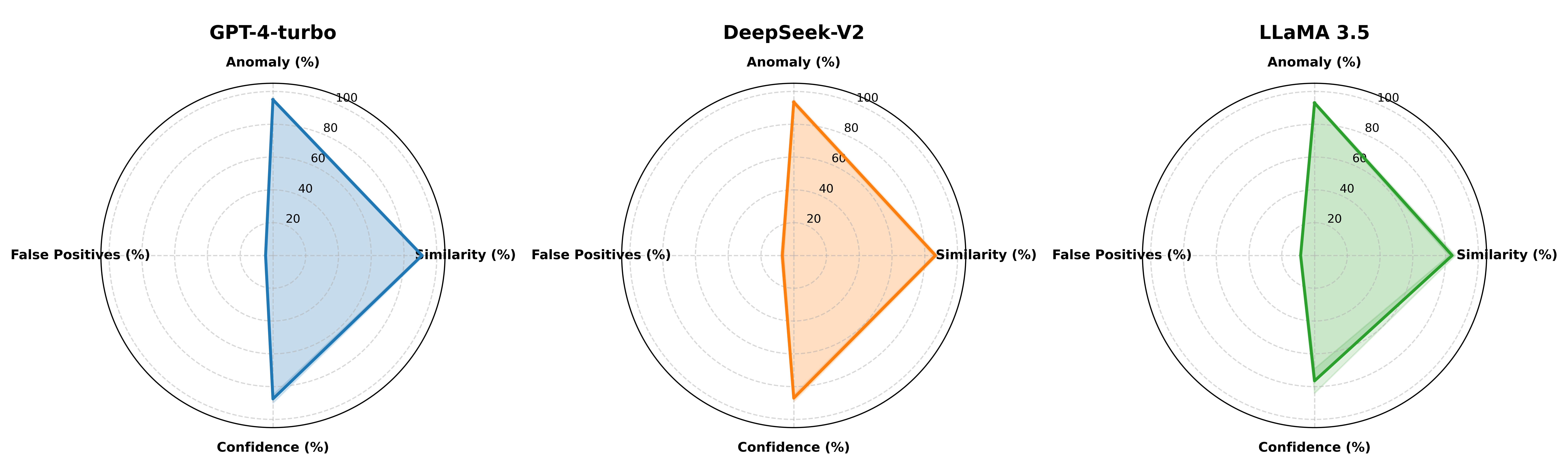}
    \caption{Security and reasoning quality under DoS.}
    \label{fig:dos_quality}
\end{figure*}

GPT-4-turbo consistently outperformed the other LLMs in reasoning quality; specifically, it demonstrated strong multi-metric reasoning by aligning anomalies (94-96\%), bandwidth usage (92-110 MB/s), and latency (up to 13 ms) with canonical DoS signatures. Consequently, this produced similarity scores of 90-92\% while maintaining the lowest false positive risk of 4-5\%. Moreover, GPT-4-turbo was able to cross-link telemetry, including CPU, bandwidth, and latency, and generate contextualized mitigation strategies such as SYN cookies, WAF rate-limiting, and IP blocking. Furthermore, these recommendations demonstrate a layered defense and highlight its ability to transition from anomaly detection to actionable cyber defense, as shown in \autoref{fig:dos_quality}.\\  
By contrast, DeepSeek-V2 performed reliably in terms of numeric accuracy, achieving similarity levels of 85-88\% and confidence scores of 85-89\%. However, its reasoning was more rigid and template-driven, since the model frequently emphasized CPU usage (85-89\%) and anomaly scores (92-95\%) while neglecting secondary telemetry such as bandwidth and memory. As a result, this narrow focus reduced interpretability and limited causal reasoning. Although detection itself remained correct, its explanations were repetitive and lacked cross-metric integration, which is important for diagnosing complex DoS variants that exploit multiple system bottlenecks. Consequently, false positive risk was higher (6-8\%).\\  
On the other hand, LLaMA3.5 consistently lagged, achieving only 82-86\% similarity and 69-84\% confidence, with false positive risk reaching 8-9\%. In particular, its reasoning was shallow, as it often focused on single indicators such as latency spikes and memory usage above \SI{1300}{\mega\byte}. This single-feature dependency made its outputs less robust and occasionally misleading, since it failed to establish cause-and-effect links between anomalous metrics. In some instances, attacks were flagged only as ``warnings,'' which could delay mitigation. For a mission-critical edge gateway, this weakness makes LLaMA3.5 unsuitable as a primary detection engine. 

\subsubsection{Resource Consumption}
\begin{figure*}[htbp]
    \centering
    \includegraphics[width=0.9\linewidth]{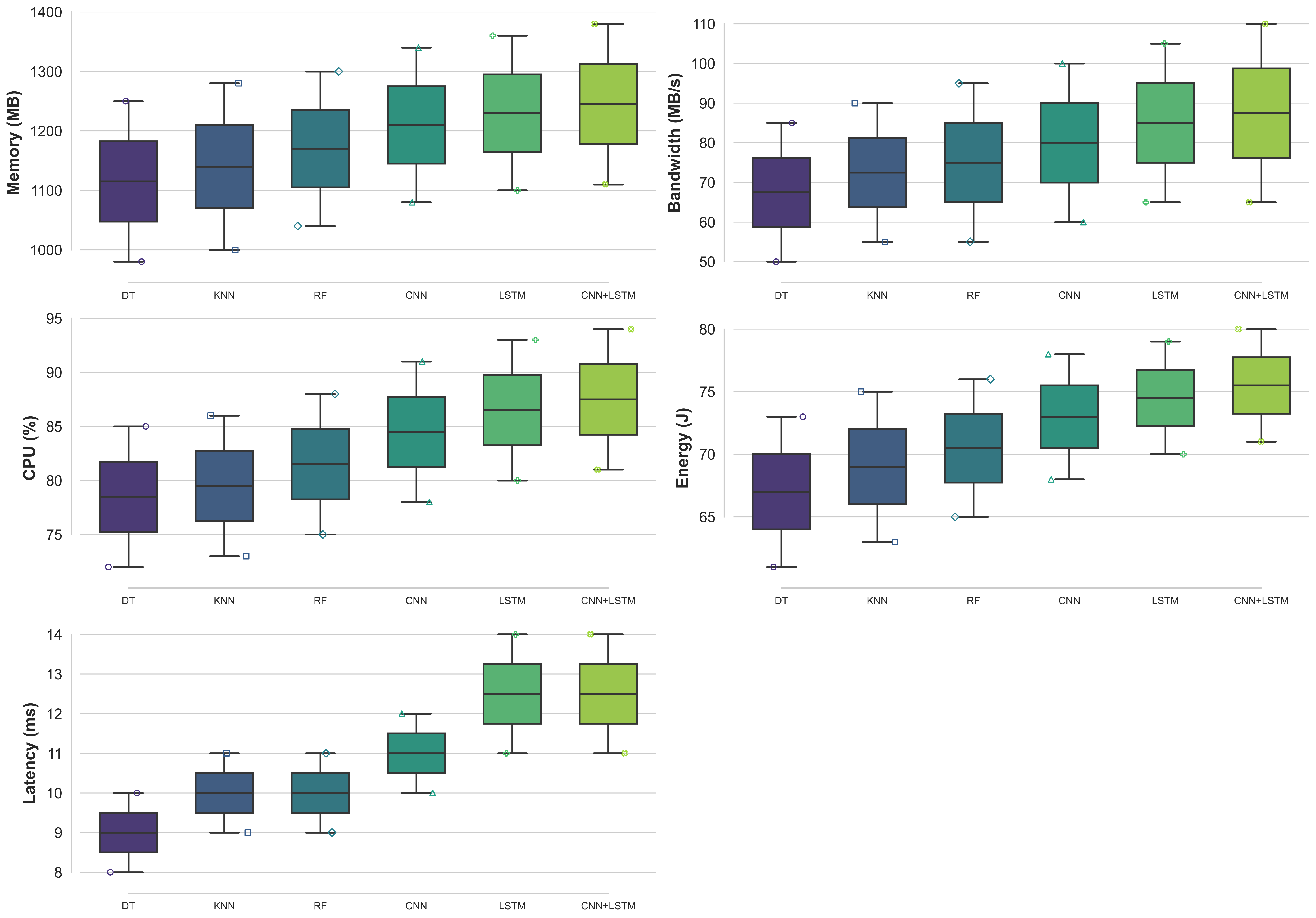}
    \caption{Resource usage under DoS.}
    \label{fig:dos_resources}
\end{figure*}

Resource consumption analysis revealed a clear trend: deeper IDS models (CNN, LSTM, and a hybrid model of CNN and LSTM) demanded significantly higher resources under DoS, whereas classical ML-based IDS models (DT, KNN, RF) remained lightweight and efficient. Specifically, memory usage spanned from 980 MB to 1380 MB, with LLaMA3.5-based configurations tending toward the higher end. CPU usage ranged from 72-94\%, thereby reflecting the strain of sustained DoS. Moreover, bandwidth usage varied between 50 and 110 MB/s, highlighting the network stress imposed by flooding attempts. Energy demand ranged from 61 J to 80 J, and latency values spanned 8-14 ms. These patterns are summarized in \autoref{fig:dos_resources}.\\  
Classical ML-based IDS models, e.g., DT, KNN, and RF, have proven suitable for edge gateway, as efficiency is critical in constrained environments. Their lower memory and CPU requirements allowed them to sustain operations while still enabling semantic reasoning through LLM integration. Moreover, by contrast, CNN and LSTM-based architectures consumed significantly more resources, a trade-off that could limit their applicability at the edge. Consequently, these findings suggest that LLM choice has little impact on raw resource usage, whereas ML-based IDS model complexity remains the primary determinant of operational cost.

\subsubsection{Statistical Analysis}
To determine the statistical significance of these observations, ANOVA and Tukey HSD tests were applied across all LLM-integrated ML-based IDS models. As summarized in \autoref{tab:dos_stats}, the results confirmed that memory usage was the only metric with statistically significant differences ($p < 0.0001$), with LLaMA3.5 consuming substantially more than GPT-4-turbo and DeepSeek-V2. Moreover, the effect size for memory was large ($\eta^2 = 0.949$), underscoring its practical impact. By contrast, for all other metrics, including CPU usage, bandwidth, latency, and energy, $p$-values exceeded 0.05 and effect sizes remained insignificant ($\eta^2 < 0.005$). Consequently, this indicates that LLM integration does not introduce variability in these dimensions. From a security standpoint, this consistency is advantageous, since it ensures predictable resource footprints under DoS conditions regardless of the chosen LLM.
\begin{table*}[!t]
\centering
\caption{Statistical analysis under DoS.}
\label{tab:dos_stats}
\renewcommand{\arraystretch}{1.2}
\begin{tabular}{|l|c|c|c|c|}
\hline
Metric & ANOVA F & p-value & Effect Size ($\eta^2$) & Tukey Result \\
\hline
Memory     & 9963.87 & $<$0.0001 & 0.949  & LLaMA $>$ GPT-4, DeepSeek \\
Bandwidth  & 0.045   & 0.956     & 0.00008 & No difference \\
CPU        & 0.386   & 0.680     & 0.0007  & No difference \\
Energy     & 0.385   & 0.680     & 0.0007  & No difference \\
Latency    & 1.765   & 0.172     & 0.0033  & No difference \\
\hline
\end{tabular}
\end{table*}
\begin{tcolorbox}[
  colback=green!10!white,
  colframe=green!60!black,
  coltitle=white,
  fonttitle=\bfseries,
  title=Finding,
  breakable,         
  enhanced,
  drop shadow
]
GPT-4-turbo demonstrated the strongest balance of reasoning quality, accuracy, and actionable mitigation under DoS. Its ability to integrate multiple telemetry features into coherent narratives makes it highly effective for real-time defense. DeepSeek-V2 provided numerically sound but repetitive and less insightful outputs, making it appropriate for scenarios where raw detection is prioritized over interpretability. LLaMA3.5, while functional, was limited in reasoning quality and imposed the heaviest memory burden, making it less suited for constrained deployments. Resource analysis confirmed that the choice of ML-based IDS backbone, rather than LLM, remains the dominant factor in resource usage. Thus, prioritizing semantic reasoning performance (favoring GPT-4-turbo) while deploying lightweight ML-based IDS models offers the most balanced and practical defense strategy against DoS at the edge gateway.
\end{tcolorbox}

\subsection{DDoS Attack Evaluation}
\label{sec:ddos_attack_evaluation}
This section evaluates DDoS detection with LLM and ML-based IDS integration, focusing on three core aspects: (i) security reasoning quality and interpretability of LLM outputs, (ii) system-level resource usage under large-scale distributed flooding traffic, and (iii) statistical validation of differences across models. Since DDoS attacks combine distributed origins with sustained volume, they represent one of the most resource-intensive and challenging scenarios for both ML-based IDS and LLM integration.

\subsubsection{Security and Reasoning Quality}
\begin{figure*}[htbp]
    \centering
    \includegraphics[width=0.9\linewidth]{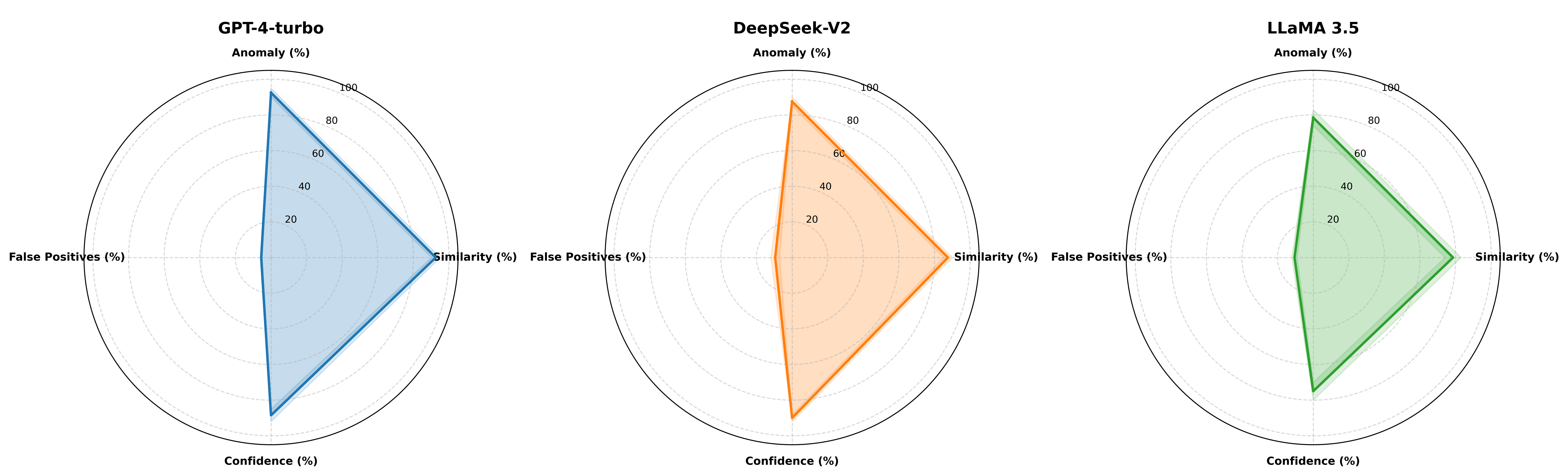}
    \caption{Security reasoning under DDoS.}
    \label{fig:ddos_quality}
\end{figure*}

GPT-4-turbo consistently provided the strongest reasoning depth, achieving 90-95\% similarity with known DDoS patterns and confidence levels of 85-92\%. In addition, its false positive rate remained at 5-6\%, which was the lowest among all LLMs. Importantly, GPT-4-turbo employed multi-step causal reasoning that linked anomaly scores, bandwidth saturation (often above 100 MB/s), and latency increases (10-13 ms) into coherent explanations. Moreover, the recommended mitigations-e.g., rate-limiting, SYN-proxying, and blackholing of distributed IP sources-highlighted its ability to generate actionable security, as shown in \autoref{fig:ddos_quality}.\\ 
By contrast, DeepSeek-V2 achieved 85-90\% similarity and 88-92\% confidence, thereby showing solid recognition accuracy. However, its reasoning style was more descriptive than analytical, since it primarily focused on CPU and memory indicators without connecting these with network-layer anomalies. Consequently, explanations often appeared formulaic and less adaptable to evolving multi-vector DDoS campaigns, even though raw classification remained correct. Furthermore, the false positives ranged between 7-12\%.\\ 
On the other hand, LLaMA3.5 lagged, with similarity restricted to 74-83\% and confidence levels between 70-80\%. Additionally, its false positive risk ranged from 9\% to 12\%. The reasoning was shallow, as it frequently highlighted isolated features, e.g., latency and CPU, without integrating them into a causal attack narrative. As a result, this fragmented reasoning increased the chance of mislabeling severe DDoS floods as moderate anomalies, thereby undermining its reliability for a mission-critical edge gateway.
\subsubsection{Resource Consumption Summary}
\begin{figure*}[htbp]
    \centering
    \includegraphics[width=0.9\linewidth]{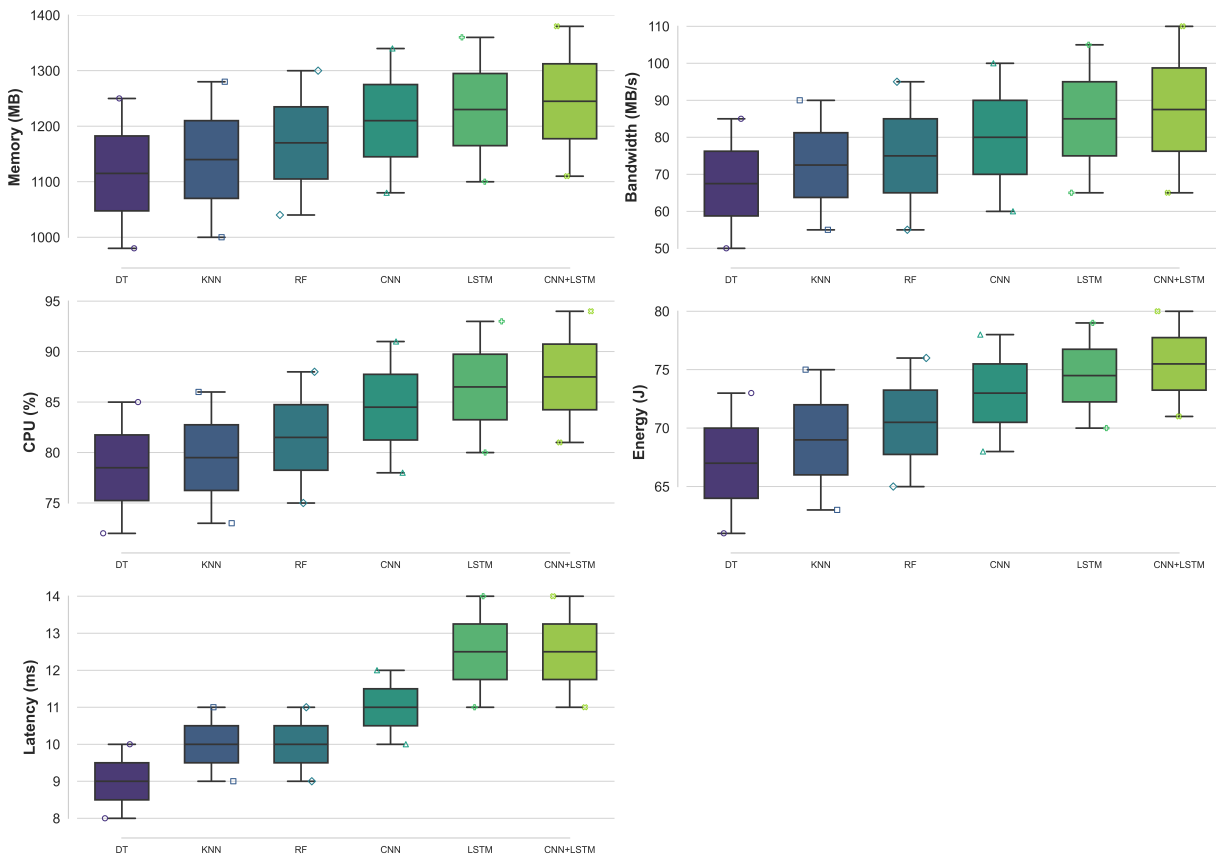}
    \caption{Resource usage under DDoS.}
    \label{fig:ddos_resources}
\end{figure*}

Resource consumption under DDoS was elevated compared to DoS because of the distributed traffic load. Specifically, memory usage ranged from 1000-1350 MB across IDS models, with CNN and LSTM variants consuming the most. Moreover, CPU usage reached 75-92\%, thereby highlighting the heavy computational demand of processing distributed flooding. In addition, energy consumption was between 65 and 80 J, and latency spanned 8-13 ms, which nevertheless remained acceptable for near real-time detection. Bandwidth stress was particularly high; however, DT, KNN, and RF managed it more efficiently, whereas CNN and LSTM models approached upper thresholds of 110 MB/s. These results are illustrated in \autoref{fig:ddos_resources}.\\ 
Taken together, these findings underscore that classical ML-based IDS models are resource-efficient, as they sustain operation with minimal degradation under large-scale traffic. By contrast, CNN and LSTM models, while offering slightly stronger detection accuracy, impose higher operational costs. Furthermore, LLM integration did not drastically alter CPU, energy, and latency footprints; nevertheless, bandwidth handling varied, reflecting GPT-4’s tendency to include richer multi-metric reasoning in its semantic outputs.
\subsubsection{Statistical Analysis}
ANOVA and Tukey HSD were applied to validate the statistical relevance of differences across metrics. The results, summarized in \autoref{tab:ddos_stats}, indicated that bandwidth was the only significantly different metric ($p<0.001$, $\eta^2=0.9725$). Moreover, Tukey analysis confirmed that GPT-4-turbo introduced slightly higher bandwidth demands compared to DeepSeek-V2 and LLaMA3.5, which is consistent with its multi-metric reasoning style that required processing larger contextual prompts. In contrast, for memory, CPU, energy, and latency, $p$-values exceeded 0.05, and effect sizes remained minimal ($\eta^2 < 0.005$). Consequently, this suggests that LLM choice does not impact these metrics in a statistically significant way. From a security perspective, this stability is desirable, since it guarantees predictable resource costs regardless of which LLM is deployed.
\begin{table*}[!t]
\centering
\caption{Statistical analysis under DDoS.}
\label{tab:ddos_stats}
\renewcommand{\arraystretch}{1.2}
\begin{tabular}{|l|c|c|c|c|}
\hline
Metric & ANOVA F & p-value & Effect Size ($\eta^2$) & Tukey Result \\
\hline
Memory     & 0.612   & 0.542 & 0.0011 & No difference \\
Bandwidth  & 19062.1 & $<$0.001 & 0.9725 & Significant difference \\
CPU        & 0.100   & 0.905 & 0.0011 & No difference \\
Energy     & 1.066   & 0.345 & 0.0020 & No difference \\
Latency    & 2.704   & 0.067 & 0.0050 & No difference \\
\hline
\end{tabular}
\end{table*}
\begin{tcolorbox}[
  colback=green!10!white,
  colframe=green!60!black,
  coltitle=white,
  fonttitle=\bfseries,
  title=Finding,
  enhanced,
   breakable,         
  drop shadow
]
GPT-4-turbo demonstrated the most effective reasoning, as it integrated multiple features into causal attack narratives and proposed layered mitigations. By contrast, DeepSeek-V2, while accurate, relied on descriptive and less adaptable outputs. Moreover, LLaMA3.5 showed fragmented reasoning and higher false positive risks, thereby limiting its use in critical environments. From a resource standpoint, classical ML-based IDS backbones proved efficient under DDoS stress, whereas deeper architectures added cost without providing significant statistical benefit. The only exception was bandwidth, since GPT-4-turbo’s reasoning demanded more resources; however, this cost is offset by its superior interpretability and actionable insights.
\end{tcolorbox}

\subsection{Port Scanning Evaluation}
\label{sec:port_attack_evaluation}
This section evaluates port scanning detection with ML-based IDS and LLM integration. Moreover, port scanning is a reconnaissance activity that generates abnormal connection surges and sequential port access, making it critical to detect early before exploitation attempts can occur. The evaluation highlights the quality of reasoning, efficiency under resource constraints, and statistical validation of observed differences across ML-based IDS and LLM settings.
\subsubsection{Security and Reasoning Quality}
\begin{figure*}[htbp]
    \centering
    \includegraphics[width=0.9\linewidth]{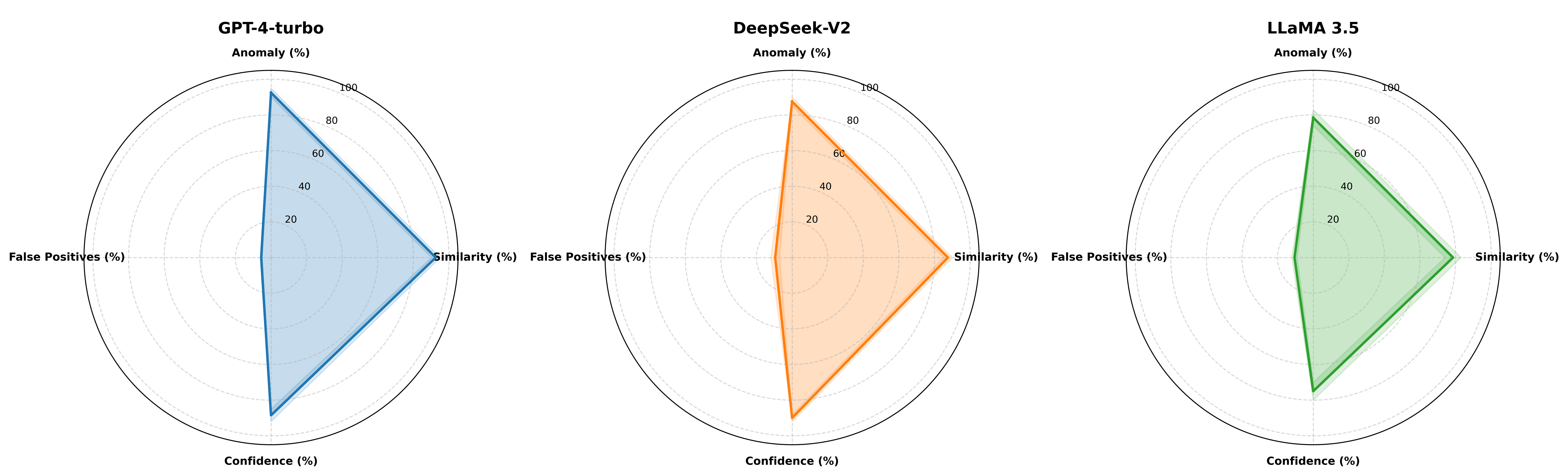}
    \caption{Security reasoning under port scanning.}
    \label{fig:port_quality}
\end{figure*}

GPT-4-turbo demonstrated the strongest interpretability, as it correlated anomaly levels (92-98\%) with connection surges and bandwidth patterns, while also factoring in latency ($\sim$11 ms). Consequently, this holistic analysis yielded similarity scores of 89-93\%, confidence of 85-92\%, and false positives around 5\%. Moreover, its causal reasoning was robust, since it narrated the relationship between rapid connection bursts and port-based probing. Recommended mitigations, therefore, included dynamic firewall rules and scan-throttling strategies, as illustrated in \autoref{fig:port_quality}.\\  
By contrast, DeepSeek-V2 achieved 83-88\% similarity and 84-90\% confidence, with false positives between 6-8\%. Although accurate, its reasoning was formulaic, as it repeatedly highlighted CPU and anomaly alignment without deeper causal connections. As a result, outputs were less useful for analysts seeking rich situational awareness.\\  
On the other hand, LLaMA3.5 remained the weakest, achieving 81-86\% similarity and 70-79\% confidence, with false positives at 8-9\%. In particular, its reasoning often isolated individual features, e.g., latency spikes, without linking them to broader scan patterns. Consequently, this reduced interpretability and occasionally led to misclassification of stealth scans as benign.\\  
\subsubsection{Resource Consumption}
\begin{figure*}[htbp]
    \centering
    \includegraphics[width=0.9\linewidth]{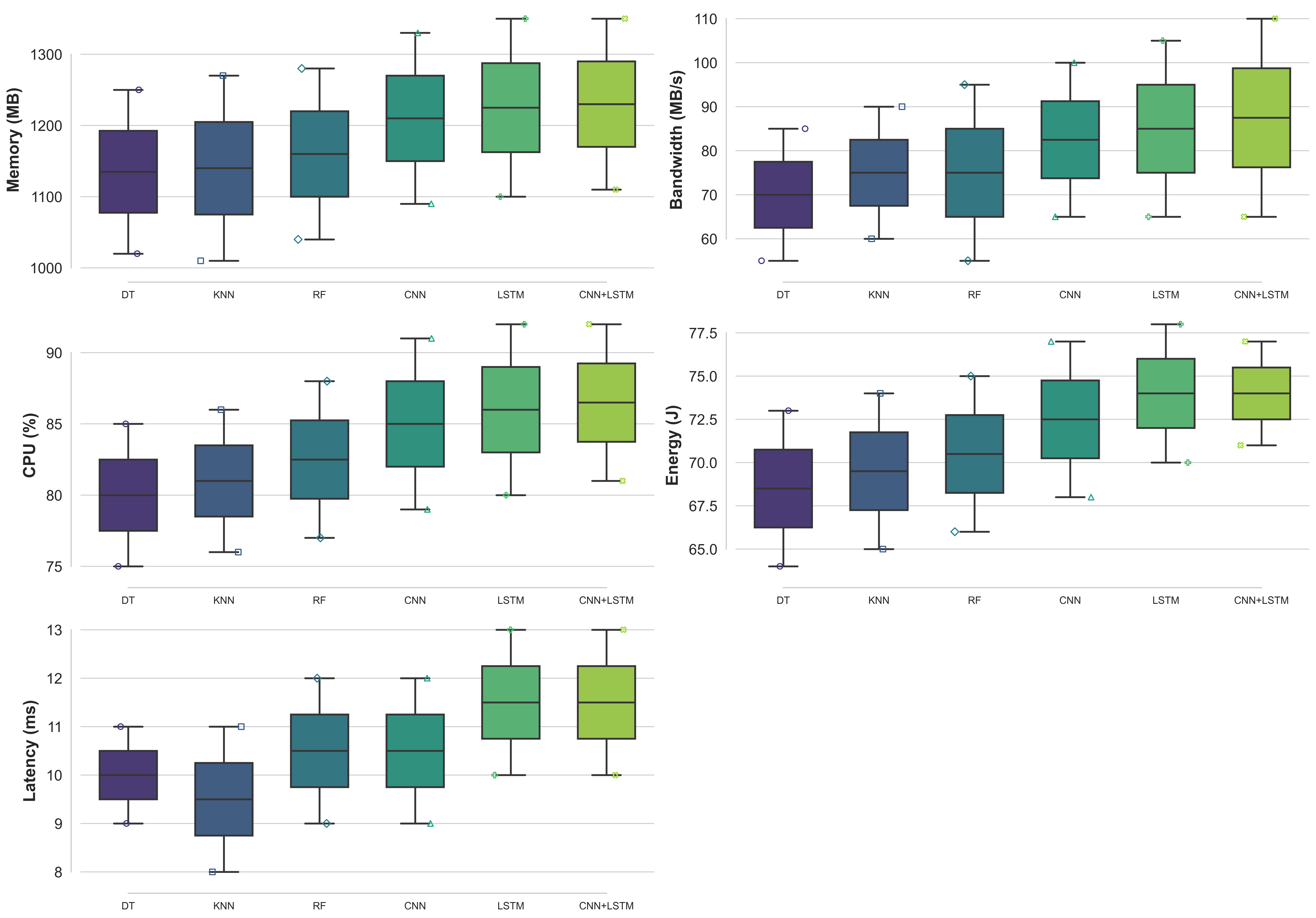}
    \caption{Resource usage under port scanning.}
    \label{fig:port_resources}
\end{figure*}

Resource consumption during port scanning remained consistent with other attack classes. Specifically, memory usage spanned 1000-1350 MB, CPU ranged from 75-92\%, and bandwidth demands were between 50-110 MB/s. In addition, energy costs varied from 63-77 J, while latency remained at 8-13 ms. Moreover, classical ML-based IDS backbones (DT, KNN, RF) sustained efficient operation, whereas CNN and LSTM consumed more resources, particularly memory and bandwidth. These results are summarized in \autoref{fig:port_resources}, thereby highlighting that despite the additional complexity of sequential scan detection, overhead nevertheless remained manageable across all ML-based IDS and LLM configurations.

\subsubsection{Statistical Analysis}
Statistical testing confirmed that no significant differences emerged across LLMs for port scanning. Specifically, the results summarized in \autoref{tab:port_stats} show that ANOVA returned $p>0.05$ for all metrics, and effect sizes were insignificant ($\eta^2 < 0.003$). Moreover, Tukey HSD further confirmed this uniformity, suggesting that the resource overhead of integrating LLMs into ML-based IDS does not vary meaningfully under scan detection. Consequently, this stability is valuable for deployment, since it guarantees predictable performance across heterogeneous edge gateways.
\begin{table*}[!t]
\centering
\caption{Statistical analysis under port scanning.}
\label{tab:port_stats}
\renewcommand{\arraystretch}{1.2}
\begin{tabular}{|l|c|c|c|c|}
\hline
Metric & ANOVA F & p-value & Effect Size ($\eta^2$) & Tukey Result \\
\hline
Memory     & 0.256 & 0.774 & 0.0005 & No difference \\
Bandwidth  & 0.037 & 0.964 & 0.0004 & No difference \\
CPU        & 0.901 & 0.407 & 0.0017 & No difference \\
Energy     & 0.901 & 0.407 & 0.0017 & No difference \\
Latency    & 1.228 & 0.293 & 0.0065 & No difference \\
\hline
\end{tabular}
\end{table*}
\begin{tcolorbox}[
  colback=green!10!white,
  colframe=green!60!black,
  coltitle=white,
  fonttitle=\bfseries,
  title=Finding,
   breakable,         
  enhanced,
  drop shadow
]
GPT-4-turbo delivered the most interpretable port scanning detection, as it generated coherent, causal explanations linking anomaly scores, latency, and connection surges. Furthermore, by contrast, DeepSeek-V2 offered accurate detection but relied on repetitive and narrow reasoning, whereas LLaMA3.5 was shallow and inconsistent, often isolating single features without integration. Moreover, since resource usage patterns were stable across all models, the decisive factor remains reasoning quality, thereby making GPT-4-turbo the most effective choice for early reconnaissance threat detection.
\end{tcolorbox}

\section{Cross-Attacks Analysis}
\label{sec:cross_attack_discussion}
This section addresses RQ3, as it synthesizes the results across four attack categories( Brute Force, DoS, DDoS, and Port Scanning) by comparing mL-based IDS and LLM hybrid performance. Specifically, the comparative discussion emphasizes three axes: (i) reasoning quality and interpretability of LLM outputs, (ii) resource consumption patterns across ML-based IDS and LLM configurations, and (iii) statistical validation of observed differences. Taken together, these findings provide a holistic understanding of how semantic reasoning impacts the robustness of ML-based IDS in a real edge gateway, and consequently, they offer direct evidence in support of answering RQ3.

\subsection{Security and Reasoning Quality Across Attacks}
\begin{figure*}[htbp]
    \centering
    \includegraphics[width=0.9\linewidth]{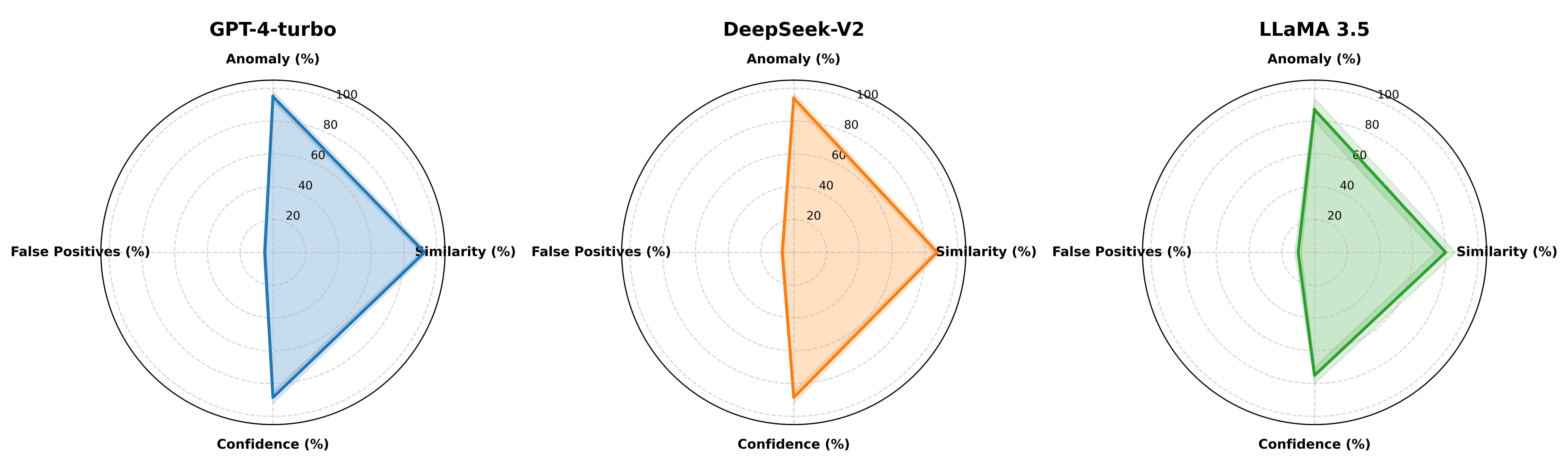}
    \caption{Comparative security reasoning performance of LLMs across attack categories.}
    \label{fig:all_quality}
\end{figure*}

GPT-4-turbo consistently demonstrated the most coherent and actionable reasoning across all four attack categories. Specifically, it achieved anomaly alignment of 92-98\%, similarity scores above 90\%, semantic confidence in the 85-92\% range, and maintained the lowest false positive risk (4-6\%). Moreover, its reasoning combined multiple telemetry dimensions, e.g., anomaly levels, CPU usage, bandwidth surges, memory usage, and latency variations, into coherent causal narratives. Consequently, security recommendations, i.e.,  MFA, IP blocking, SYN cookies, and WAF throttling, were context-specific, thereby making GPT-4 outputs both accurate and operationally useful, as shown in \autoref{fig:all_quality}.\\  
By contrast, DeepSeek-V2 also provided high anomaly accuracy (91-97\%), with similarity in the 85-90\% range and semantic confidence between 85-92\%. However, its interpretability lagged because of formulaic reasoning, as explanations often emphasized a single metric (e.g., CPU or anomaly rate) while neglecting secondary features such as bandwidth and energy consumption. As a result, while DeepSeek-V2 was reliable in raw detection, its explanatory depth remained limited, which hindered analysts during forensic investigations or layered defense planning.\\  
On the other hand, LLaMA3.5 consistently lagged in semantic integration. Its anomaly scores ranged between 80-94\%, similarity dropped to 74-86\%, confidence fell to 70-80\%, and false positives were higher (8-12\%). In particular, its reasoning often isolated one or two signals (e.g., memory overhead in DoS, latency spikes in DDoS) without integrating them into a coherent story. Consequently, this caused critical intrusions to be occasionally downgraded to ``warnings,'' thereby representing a potential blind spot in the mission-critical edge gateway. 
\subsection{Resource Consumption Comparison}
\begin{figure*}[htbp]
    \centering
    \includegraphics[width=0.9\linewidth]{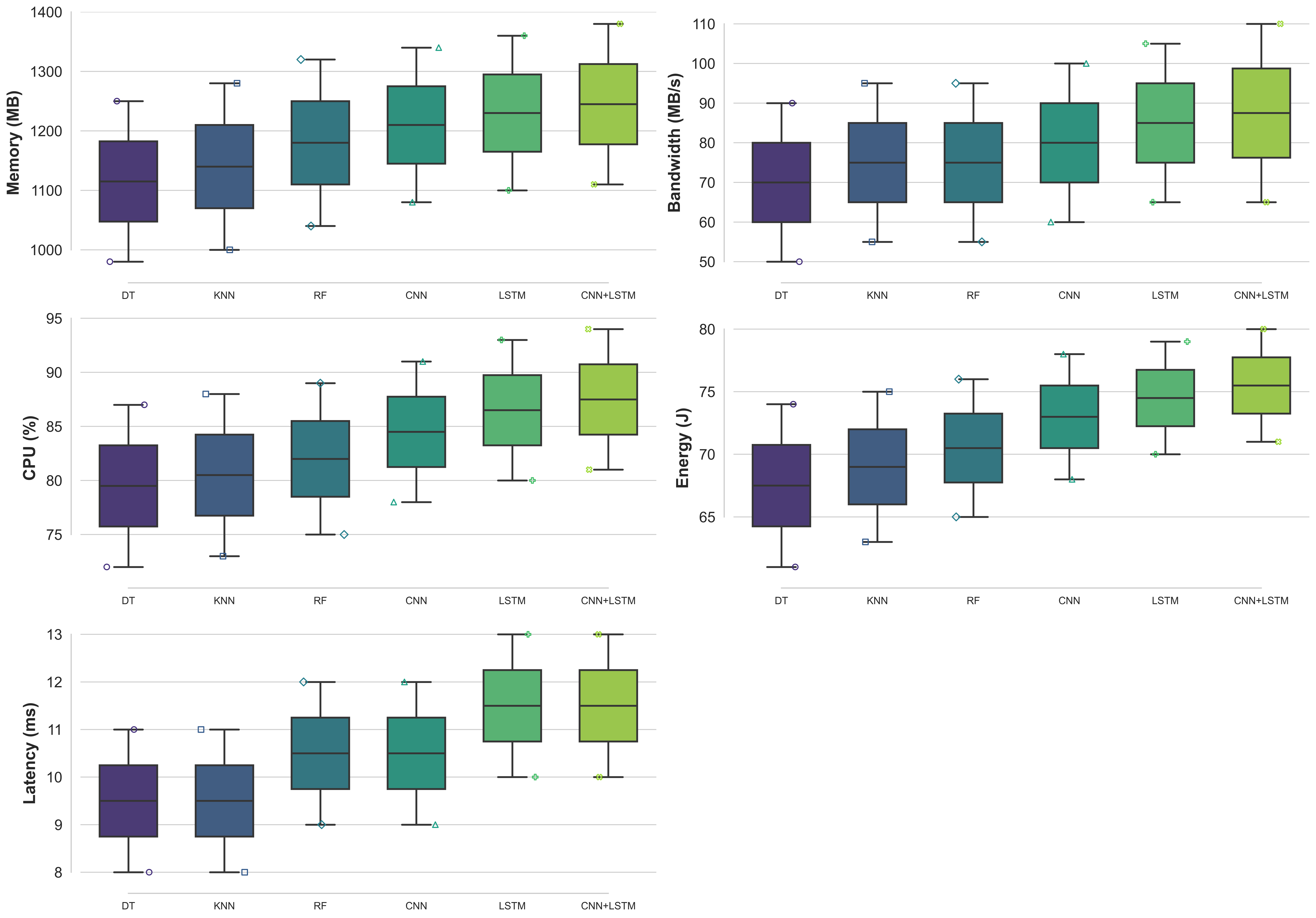}
    \caption{Resource consumption across ML-based IDS and LLM hybrids under all attack categories.}
    \label{fig:all_resources}
\end{figure*}

Resource costs were broadly stable across all attack types, thereby confirming that IDS-LLM integration introduces minimal additional overhead. Specifically, the typical resource ranges observed included 950--1380 MB of memory, 72-95\% CPU utilization, bandwidth consumption of less than 120 MB/s, energy draw between 61-80 J, and latency of 8-14 ms. Moreover, these values remained consistent across Brute Force, DoS, DDoS, and Port Scanning scenarios, as summarized in \autoref{fig:all_resources}.\\  
In terms of model efficiency, classical IDS models (DT, KNN, RF) were consistently more efficient, as they consumed fewer resources while still maintaining adequate detection accuracy. By contrast, CNN, LSTM, and the hybrid model of CNN and LSTM incurred higher memory and CPU costs, yet they achieved stronger raw detection capability. Consequently, this trade-off reinforces the suitability of lightweight models for constrained IoT gateways, whereas deep models remain valuable in deployments where stronger detection is prioritized over efficiency.\\  
Among the LLMs, GPT-4-turbo and DeepSeek-V2 showed similar resource profiles. However, LLaMA3.5 consumed significantly more memory under DoS conditions. A second notable trend emerged under DDoS, where GPT-4-turbo introduced higher bandwidth peaks due to larger reasoning outputs; nevertheless, these remained within operational IoT thresholds.

\subsection{Statistical Analysis}
One-way ANOVA with Tukey HSD post-hoc testing revealed that for most metrics, LLM differences were statistically insignificant ($p > 0.05$, $\eta^2 < 0.005$), as summarized in \autoref{tab:all_stats}. This result supports the conclusion that LLM integration does not substantially alter system efficiency under most conditions. However, two significant exceptions emerged.\\
First, under DoS conditions, LLaMA3.5 consumed substantially more memory, with results highly significant ($p < 0.0001$, $\eta^2 \approx 0.95$). This makes LLaMA3.5 unsuitable for constrained IoT environments where memory overhead is a critical concern. Second, during DDoS scenarios, GPT-4-turbo exhibited consistently higher bandwidth demands, significant at $p < 0.001$, $\eta^2 \approx 0.97$. Although these demands remained within edge tolerances, they suggest that bandwidth-aware policies require stable deployment.
\begin{table*}[!t]
\centering
\caption{Aggregate statistical analysis across attack categories.}
\label{tab:all_stats}
\renewcommand{\arraystretch}{1.2}
\begin{tabular}{|l|c|c|c|c|}
\hline
Metric (by Attack) & ANOVA F & p-value & Effect Size ($\eta^2$) & Tukey Result \\
\hline
DoS Memory     & 9963.87  & $<$0.0001 & 0.949 & LLaMA $>$ GPT-4, DeepSeek \\
DDoS Bandwidth & 19062.1  & $<$0.001  & 0.9725 & GPT-4 $>$ DeepSeek, LLaMA \\
CPU (all)      & 0.386    & 0.680     & 0.0007 & No difference \\
Energy (all)   & 0.901    & 0.407     & 0.0017 & No difference \\
Latency (all)  & 1.228    & 0.293     & 0.0065 & No difference \\
\hline
\end{tabular}
\end{table*}

\begin{tcolorbox}[
  colback=green!10!white,
  colframe=green!60!black,
  coltitle=white,
  fonttitle=\bfseries,
  title=Finding,
  enhanced,
   breakable,         
  drop shadow
]
GPT-4-turbo emerged as the most reliable LLM for ML-based IDS integration. Specifically, it consistently achieved anomaly alignment above 92\%, similarity greater than 90\%, minimized false positives, and delivered actionable, multi-metric reasoning. By contrast, DeepSeek-V2 struck a balance between accuracy and efficiency but was less interpretable, thereby making it more suitable in contexts where numeric precision outweighs explainability. On the other hand, LLaMA3.5, while functional, was resource-heavy under DoS and semantically shallow across all attacks, which makes it better suited as a backup option rather than a primary detection engine. From the ML-based IDS backbone perspective, DT, KNN, and RF remain optimal for constrained devices. In contrast, CNN, LSTM, and a hybrid of CNN and LSTM enhance detection robustness but at a higher computational cost. Ultimately, these insights confirm that the quality of security reasoning, rather than raw computational overhead, should guide the selection of LLMs at the edge gateway.
\end{tcolorbox}

\section{Runtime Log Example}
\label{sec:log_1}
To illustrate the entire operation of the proposed edge-based semantic ML-based IDS, Figure~\ref{fig:runtime_log} presents a structured runtime log generated during a brute-force attack. It demonstrates the multi-model anomaly detection process, real-time system telemetry, context-based prompt creation, LLM-based enhanced, and mitigation feedback.
\begin{figure*}[ht]
\centering
\begin{minipage}{0.95\textwidth}
\begin{lstlisting}[style=logstyle, label={lst:runtime_log}, escapeinside={(*@}{@*)}]
[Edge Node: RPi-Gateway-01]  Timestamp: 2025-06-25 14:08:46
==================================================================
-> New traffic window detected [session_id: 7492]
-> Feature vector extracted:
   x_t = [0.88, 16, 4.2, 0, 3389, 22, 0.61, 0.97, 230, 18, 0.07, 0.54]

-> Running predictions across ML-based IDS models...

[*] DT                         -> Label: brute force   | Score: 0.91
[*] KNN                        -> Label: brute force   | Score: 0.89
[*] RF                         -> Label: brute force   | Score: 0.94
[*] LSTM                       -> Label: brute force   | Score: 0.96
[*] CNN                        -> Label: brute force   | Score: 0.92
[*] Hybrid CNN and LSTM        -> Label: brute force   | Score: 0.97

(*@\textbf{\textcolor{orange}{-> Consensus: 6/6 models classify as "brute force"}}@*)
(*@\textbf{\textcolor{red}{-> Aggregated anomaly score s\_t = 0.93 >= tau\_alert = 0.70 -> ALERT triggered}}@*)

-> System metrics captured:
   CPU = 47.6\%, Memory = 372 MB, Latency = 48.2 ms, Energy = 21.7 J
   Normalized telemetry vector:
   m_t = [0.476, 0.182, 0.964, 0.072, 0.930]

-> Retrieved context for class "brute force":
   "Repeated login attempts over network protocols such as SSH or RDP,
    Typically using dictionary-based or credential-stuffing attacks."

-> Constructing LLM prompt with telemetry and context...
-> Sending prompt to external LLM: GPT-4 Turbo

(*@\textbf{\textcolor{blue}{=> LLM Response [elapsed: 0.84 sec]}}@*)
{
  "revised_label": "brute force",
  "confidence": 0.95,
  (*@\textbf{\textcolor{red}{"severity": "Critical",}}@*)
  "mitigation": [
    "Block the offending IP address",
    "Apply rate limiting on authentication endpoints",
    "Enforce multi-factor authentication."
  ]
}

(*@\textbf{\textcolor{cyan}{-> Final enriched classification:}}@*)
   - Class      : brute force
   - Confidence : 95\%
   (*@\textbf{\textcolor{red}{- Severity   : Critical}}@*)
   (*@\textbf{\textcolor{violet}{- Mitigation : IP block, login throttling, and enforce multi-factor authentication}}@*)

-> Total round-trip latency: 1.32 s
-> Total energy consumption: 23.9 J
(*@\textbf{\textcolor{green}{-> Mitigation instructions have been dispatched to the local firewall.}}@*)
==================================================================
\end{lstlisting}
\end{minipage}
\caption{Log showing ML-based IDS, LLM, and automated mitigation pipeline on the edge gateway.}
\label{fig:runtime_log}
\end{figure*}

\section{Analysis of LLM-Integrated ML-based IDS Performance}
This section presents a detailed analysis of the performance, resource overhead, and qualitative benefits of integrating LLMs and  ML-based IDS. Through quantitative metrics, we evaluate the trade-offs introduced by LLM semantic improvement across latency, energy consumption, detection performance, interpretability, and reasoning modes.

\subsection{Latency and Energy Impact}
\begin{figure*}[h]
    \centering
    \includegraphics[width=0.75\linewidth]{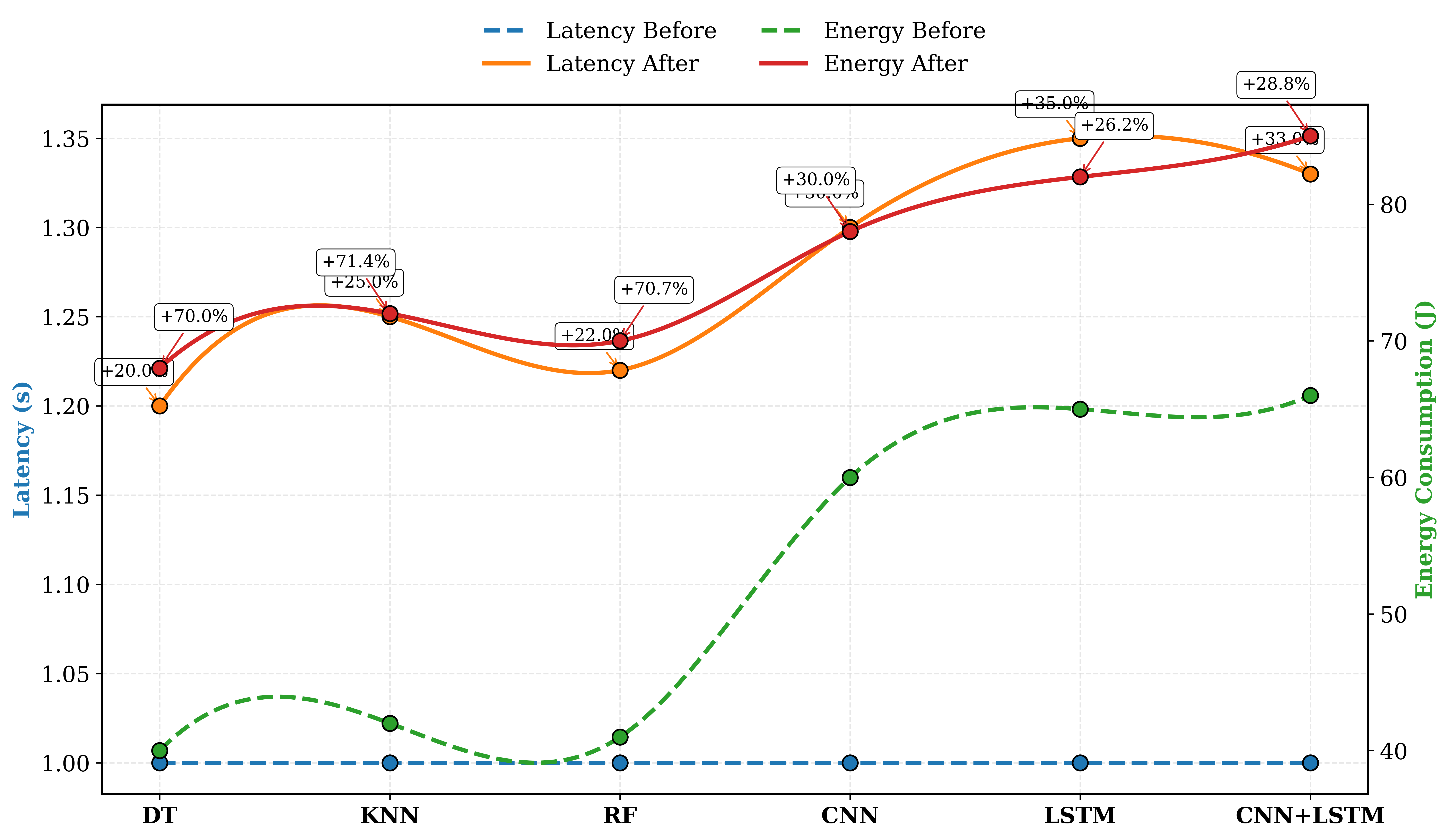}
    \caption{Latency and energy consumption comparison before and after LLM integration across ML-based IDS}
    \label{fig:latency_energy_increase}
\end{figure*}

Figure~\ref{fig:latency_energy_increase} illustrates the latency and energy consumption before and after incorporating LLMs across various ML-based IDS. The latency increment remains under 35\% across all models, ensuring suitability for real-time operations. Especially, simpler models such as DT and RF experienced lower latency and energy increases compared to deeper models, e.g,  CNN and LSTM, which showed the highest resource demand (28.8\% latency increase, 33\% energy rise). Despite these increases, all models stay within acceptable thresholds ($\leq$1.5s latency and $\leq$100J energy), validating the deployment feasibility of our framework in the edge gateway.
\subsection{Detection Performance Improvement}
\begin{figure*}[h]
    \centering
    \includegraphics[width=0.75\linewidth]{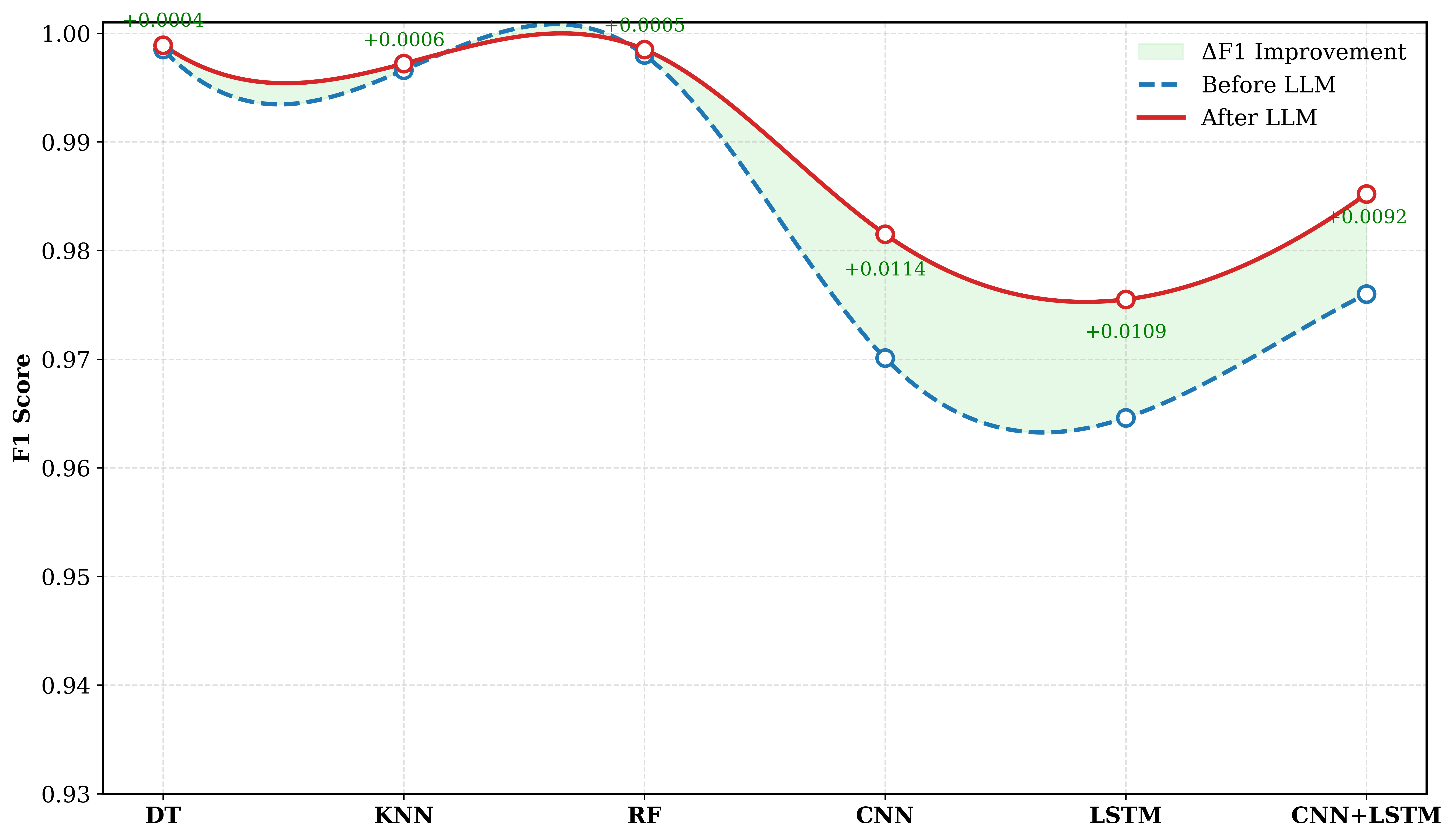}
    \caption{Impact of semantics on F1-Score improvement across ML-based IDS}
    \label{fig:semantic_enrichment_impact}
\end{figure*}

Figure~\ref{fig:semantic_enrichment_impact} demonstrates the improvement in F1-Score ($\Delta$F1) for each ML-based IDS post-LLM. Although all models demonstrate performance improvements, DL-based IDS, such as the hybrid model of CNN and LSTM, benefit the most, with F1 improvements of 0.0114 and 0.0109, respectively. These results confirm the effectiveness of semantic reasoning in improving detection accuracy, particularly for complex models that process intricate data patterns.
\subsection{Interpretability and Reasoning Mode Latency}
\label{Interpretability and Reasoning Mode Latency}
\begin{figure*}[ht]
    \centering
    \includegraphics[width=0.8\linewidth]{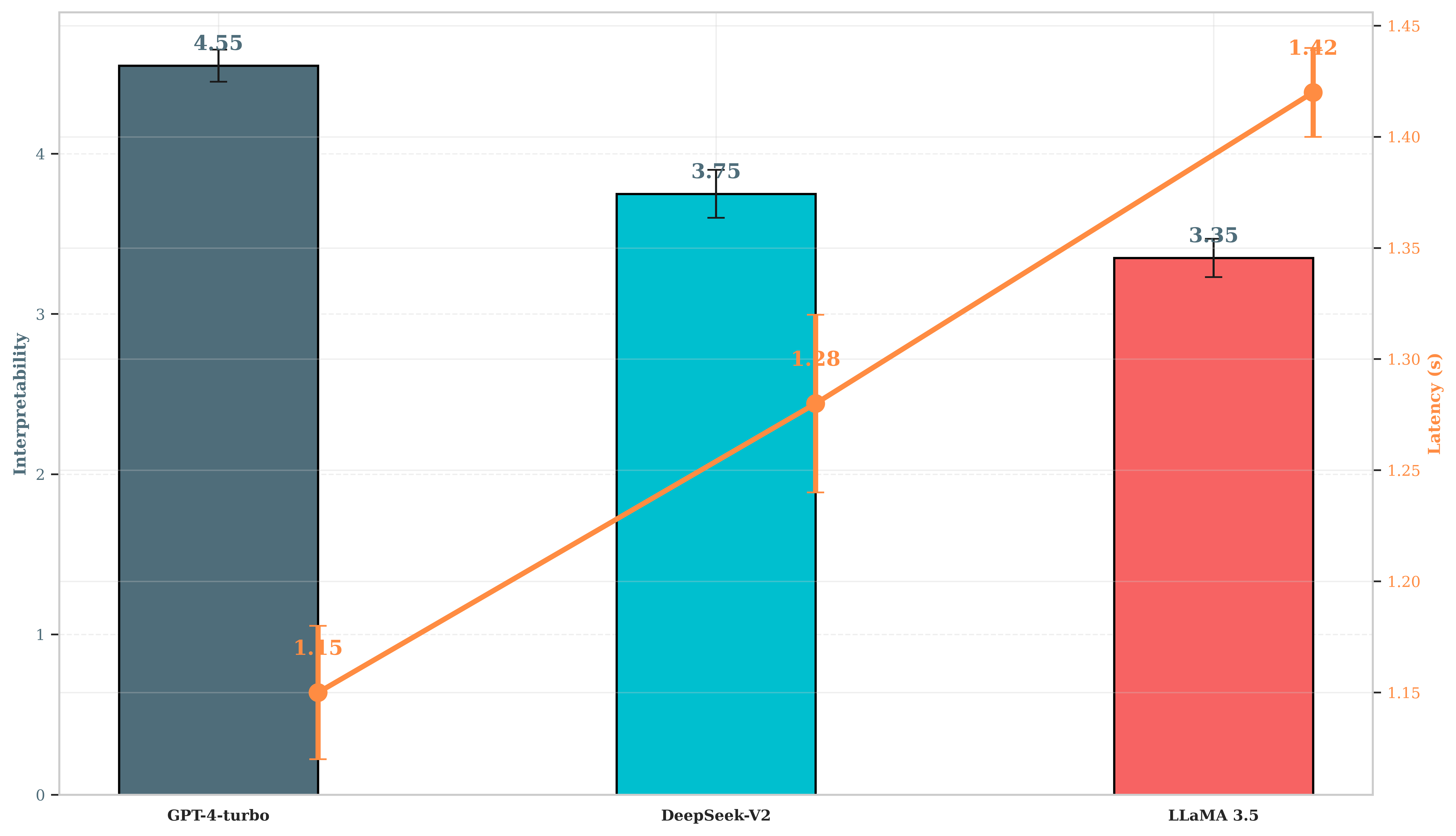}
    \caption{Comparison of interpretability and latency across GPT-4-turbo, DeepSeek-V2, and LLaMA 3.5.}
    \label{fig:interpretability_latency}
\end{figure*}

Figure~\ref{fig:interpretability_latency} summarizes the trade-offs between interpretability and latency across GPT-4-turbo, DeepSeek-V2, and LLaMA 3.5. GPT-4-turbo achieves the highest interpretability \( \text{median} > 4.5 \), essential for scenarios requiring human-in-the-loop validation. Conversely, latency varies with reasoning modes: zero-shot achieves the fastest response (~1.15s), while CoT incurs the highest latency (~1.45s) due to deeper reasoning. 

\subsection{Model Performance Evaluation}
We benchmark our proposed LLM-enhanced IDS framework against several recent state-of-the-art ML-based IDS developed for the edge gateway. As summarized in \autoref{tab:comparison}, \textit{SecurityBERT}~\cite{ferrag2024revolutionizing} uses a 15-layer BERT with privacy-preserving encoding and achieves 98.2\% accuracy with low latency (<0.15s) and compact size (16.7MB), though it lacks semantic reasoning capabilities. \textit{Fed-BERT}~\cite{adjewa2024efficient} reaches 97.79\% accuracy in centralized and federated modes with latency around 0.45s, supporting edge deployment via quantization. The \textit{TL-CBAM-EL} model~\cite{abdelhamid2024attention} employs attention and transfer learning, achieving 99.93\% accuracy; however, it requires image transformation, which limits its real-time applicability. \textit{ETree, RF and DNN}~\cite{aldaej2024ensemble} applies a two-stage ensemble achieving 98.5\% accuracy, though lacking latency and energy benchmarks. Lastly, the \textit{GPT embedding-based approach}~\cite{nwaforevaluating} achieves competitive accuracy but is computationally intensive and unsuitable for the edge gateway. In contrast, our framework delivers balanced performance across latency, interpretability, and energy, making it well-suited for the edge gateway.
\begin{table*}[ht]
\centering
\caption{Comparison of Our Framework with State-of-the-Art IDS Models}
\label{tab:comparison}
\renewcommand{\arraystretch}{1.2}
\setlength{\tabcolsep}{5pt}
\small
\resizebox{\textwidth}{!}{%
\begin{tabular}{|l|c|c|c|c|c|c|}
\hline
\textbf{Metric} & \textbf{Our Framework} & \textbf{SecurityBERT~\cite{ferrag2024revolutionizing}} & \textbf{Fed-BERT~\cite{adjewa2024efficient}} & \textbf{TL-CBAM-EL~\cite{abdelhamid2024attention}} & \textbf{ETree+RF+DNN~\cite{aldaej2024ensemble}} & \textbf{GPT Embed~\cite{nwaforevaluating}} \\
\hline
Accuracy (Top Model) & 98.9\% (CNN+LSTM) & 98.2\% & 97.79\% & 99.93\% & 98.5\% (avg) & $\sim$98\% \\
\hline
Latency & $<$1.5s (LLM+ML) & $<$0.15s & 0.45s & High & Not reported & Not reported \\
\hline
Energy Consumption & $<$100J & Low & Medium & High & Medium & High \\
\hline
Model Size & $\sim$20MB & 16.7MB & $\sim$25MB & $>$50MB & Medium & $>$300MB \\
\hline
Interpretability & High (LLM-CoT) & Moderate & Limited & Low & None & Moderate \\
\hline
Edge Suitability & High & High & High & Low & High & Low \\
\hline
\end{tabular}%
}
\end{table*}

\section{Discussion}
\label{Discussion}
This study aimed to design a resource-aware ML-based IDS framework tailored explicitly for edge gateways. By combining fast, ML-based anomaly detection with modular edge deployment, we aimed to achieve real-time responsiveness with minimal system burden. The results strongly suggest that this approach is both technically sound and practically viable. One of the most significant findings is that the tested ML-based IDS demonstrated consistent and stable performance across key resource metrics. During port scanning, statistical analyses showed no significant differences in bandwidth, latency, CPU usage, energy consumption, and memory usage across deployments using different LLM integrations. In other words, despite differences in ML-based IDS, the impact on edge system performance remains insignificant. This provides system designers with the confidence to implement such models without compromising real-time efficiency and resource availability. Furthermore, when benchmarked against state-of-the-art IDS architectures, e.g., \textit{SecurityBERT}~\cite{ferrag2024revolutionizing}, \textit{Fed-BERT}~\cite{adjewa2024efficient}, and ensemble-based approaches like \textit{ETree+RF+DNN}~\cite{aldaej2024ensemble}, the proposed system performs competitively. Although some alternatives demonstrate marginal gains in detection accuracy, they often require heavier on-device computation, increased energy demands, and lack the lightweight adaptability necessary at the edge.\\
In contrast, our method offers a balanced trade-off: accurate detection, rapid inference, and minimal overhead. Some challenges remain. The current implementation utilizes static ML pipelines and predefined detection parameters, which perform well against known attack types but require adaptation in the presence of zero-day exploits and evolving threat signatures. Future improvements include dynamic model retraining, online learning modules, and integration with decentralized learning strategies (e.g., federated learning) to accommodate concept drift. Furthermore, while the inclusion of LLMs significantly enhances interpretability and actionable threat intelligence, this benefit comes with trade-offs in energy consumption and reliance on external API services. This dependency is unsuitable for highly constrained and offline environments. Exploring on-device distilled LLMs and efficient quantized models addresses these concerns. Additionally, while we demonstrated minimal latency overhead ($<$1.5 s), time-critical healthcare systems require even faster semantic reasoning. Optimizing the LLM prompt construction process and integrating retrieval-augmented mechanisms with low-latency knowledge bases further improve performance.

\section{Threats to Validity}
\label{sec:T_V}
Empirical research in cyber-physical systems, particularly in IDS for IoT, faces inherent threats to validity. Following the guidelines of Wohlin et al.~\cite{wohlin2012experimentation}, this section outlines limitations and the mitigation steps undertaken.
\subsection{Internal Validity}
The internal validity of our results depends on the accuracy of resource measurements (e.g., CPU usage, memory utilization, latency, and energy consumption) as well as the correctness of the ML-based IDS process. To minimize measurement errors, all experiments were conducted under controlled and repeatable testbed conditions on Raspberry Pi 4 edge gateways. Each configuration (LLM and ML-based IDS) was evaluated across 62 independent trials to reduce the influence of transient system or network variability. System telemetry was captured using validated logging tools, and energy consumption was measured with standardized watt-meter instrumentation. LLM prompts were programmatically generated in a fixed format with structured reasoning instructions to ensure response consistency. Results were then averaged across trials and subjected to robust statistical analysis (ANOVA and Tukey HSD), which allowed us to control random variation and improve confidence in the observed effects.
\subsection{External Validity}
Although the results demonstrate promise for real-time LLM-enhanced ML-based IDS on the edge gateways, generalization beyond our testbed is subject to external threats. These include hardware variability, network topology differences, and the diversity of LLMs available for integration. Experiments targeted representative cyberattacks (brute force, port scanning, DoS, and DDoS) using the CICIDS2017 dataset and simulated traffic. Moreover, CICIDS2017 is not explicitly IoT-centric; it remains one of the most comprehensive, publicly available, and widely adopted datasets for intrusion detection research. It captures modern attack behaviors at the network flow level, which are directly applicable to IoT networks, as edge gateways also observe traffic at this same granularity. Moreover, the dataset has been extensively used in IoT security studies as a benchmark for proxying IDS approaches, enabling reproducibility and comparability across various strategies. Nonetheless, it does not capture all emerging IoT-specific attacks (e.g., protocol-level exploits or sensor spoofing), and this represents a limitation of the study. Reliance on externally hosted LLMs also introduces operational risks related to network availability, latency variability, and privacy, since prompts must be transmitted outside the local edge environment. However, our modular architecture allows for the substitution of different LLM providers or the deployment of lightweight on-device models with minimal reconfiguration. Similarly, while evaluations were performed on Raspberry Pi 4 platforms, performance, latency, and energy profiles can vary across other IoT hardware classes, including ARM-based gateways, industrial controllers, or battery-powered devices. External APIs further impose operational constraints in environments with limited connectivity or strict privacy requirements. These external factors must therefore be considered when extending the results of this study to broader IoT deployments.

\section{Future Work}
\label{Future Work}
The proposed framework demonstrates strong performance in terms of detection accuracy, resource efficiency, and semantic interpretability; however, several avenues for future research remain to be explored. First, the current implementation relies on static prompt templates and a limited set of manually curated few-shot examples to guide LLM reasoning. This approach constrains adaptability when encountering novel and evolving attack types. Future work will explore retrieval-augmented generation (RAG) techniques to dynamically improve LLM prompts with relevant context drawn from threat databases, historical attack logs, and real-time event streams. The architecture currently assumes cloud-based LLM inference, introducing vulnerabilities such as latency spikes, service interruptions, and data privacy risks. Research into on-device or edge-deployable LLMs, using quantization, pruning, and knowledge distillation, enables local fallback capabilities in bandwidth-constrained environments. 
Additionally, the system operates with a fixed pipeline for telemetry extraction and scoring. Integrating adaptive or self-learning mechanisms, such as reinforcement learning and feedback-aware prompt tuning, enables the system to evolve based on operator feedback, changes in network behavior, and emerging attack patterns.
Another direction is the incorporation of cross-device collaborative detection. While experiments focused on individual edge nodes, federated and swarm-based reasoning across multiple gateways can enhance situational awareness and detection robustness in large-scale IoT deployments. Although the evaluation utilized a representative attack dataset and real-time benchmarks, deploying the system in live production environments is essential to validate generalization, operational reliability, and user interpretability under realistic conditions. The proposed framework establishes a foundation for an extensible and intelligent IDS paradigm, supporting ongoing innovation in LLM integration, adaptive reasoning, and collaborative edge security systems.

\section{Conclusion}
\label{Conclusion}
This research introduces a practical and resource-conscious approach to IoT security by integrating ML-based IDS with external LLMs for real-time IDS and reasoning at the network edge. The proposed system achieves a delicate balance between detection accuracy, semantic interpretability, and operational efficiency, even under the constraints typical of edge gateways. By employing structured LLM reasoning strategies, zero-shot, few-shot, and CoT, the framework transforms raw anomaly detections into actionable intelligence, enhancing both automated and human-in-the-loop responses. Experimental results validate that the system operates within strict latency, bandwidth, and energy budgets while substantially improving interpretability and decision-making precision. This work not only advances the state of IoT security but also lays a scalable foundation for future research in deploying AI-augmented security solutions in edge gateways.

\bibliographystyle{IEEEtran}
\bibliography{cas-refs}

\end{document}